\documentclass[apj,numberedappendix,appendixfloats]{emulateapj}
\usepackage{apjfonts}
\usepackage{CJK}

\newcommand{\wise}{{\textit{WISE~}}}

\shorttitle{IR SED decomposition of Hot DOGs}
\shortauthors{L. Fan et al.}

\begin{document}

\begin{CJK*}{UTF8}{gbsn}

\title{Infrared spectral energy distribution decomposition of WISE-selected, hyperluminous hot dust-obscured galaxies}
\author{Lulu Fan (范璐璐)\altaffilmark{1,2}, Yunkun Han (韩云坤)\altaffilmark{3},  Robert Nikutta\altaffilmark{4}, Guillaume Drouart\altaffilmark{2}, and Kirsten K. Knudsen\altaffilmark{2}} \altaffiltext{1}{Shandong Provincial Key Lab of Optical Astronomy and Solar-Terrestrial Environment, Institute of Space Science, Shandong University,Weihai, 264209, China, llfan@sdu.edu.cn} \altaffiltext{2}{Department of Earth and Space Sciences, Chalmers University of Technology, Onsala Space Observatory, SE-439 92 Onsala, Sweden} \altaffiltext{3}{Yunnan Observatories, Chinese Academy of Sciences, Kunming, 650011, China, hanyk@ynao.ac.cn} \altaffiltext{4}{Instituto de Astrof\'{\i}sica, Facultad de F\'{\i}sica, Pontificia Universidad Cat\'{o}lica de Chile, 306, Santiago 22, Chile}

\begin{abstract}
  We utilize a Bayesian approach to fit the observed mid-IR-to-submm/mm spectral energy distributions (SEDs)
  of 22 \textit{WISE}-selected and submm-detected, hyperluminous hot dust-obscured galaxies (Hot DOGs),
  with spectroscopic redshift ranging from 1.7 to 4.6. We compare the Bayesian evidence of torus plus a
  gray body (Torus+GB) model with that of a torus-only (Torus) model and find that the Torus+GB model has
  the higher Bayesian evidence for all 22 Hot DOGs than the torus-only model, which represents strong evidence in favor of the Torus+GB model.
  By adopting the Torus+GB model, we decompose the observed IR SEDs of Hot DOGs into torus and cold dust components.
  The main results are: 1) Hot DOGs in our submm-detected sample are hyperluminous ($L_{IR}\geq10^{13}L_\odot$), 
  with torus emission dominating the IR energy output.
  However, cold dust emission is non-negligible, averagely contributing $\sim24\%$ of total IR luminosity. 
  2) Compared to QSO and starburst SED templates, the median SED of Hot DOGs shows the highest luminosity ratio between mid-IR and submm at rest-frame,  
  while it is very similar to that of QSOs at $\sim10-50\mu$m suggesting that the heating sources of Hot DOGs
  should be buried AGNs. 
  3) Hot DOGs have both high dust temperatures ($T_{dust}\sim72$K) and IR luminosity of cold dust.
  The $T_{dust}-L_{IR}$ relation of Hot DOGs suggests that the increase in IR luminosity for Hot DOGs is
  mostly due to the increase of the dust temperature,
  rather than dust mass. Hot DOGs have lower dust masses than those of submillimeter galaxies (SMGs) and QSOs within the similar redshift range.
  Both high IR luminosity of cold dust and relatively low dust mass in Hot DOGs can be expected by their relatively high dust temperatures.
  4) Hot DOGs have high dust covering factors, which deviate the previously proposed trend of the dust covering factor decreasing with increasing bolometric luminosity.
  Finally, we can reproduce the observed properties in Hot DOGs by employing a physical model of galaxy evolution.
  The result suggests that Hot DOGs may lie at or close to peaks of both star formation and black hole growth histories, 
  and represent a transit phase during the evolution of massive galaxies, transforming from the dusty
  starburst dominated phase to the optically bright QSO phase.

\end{abstract}


\section{Introduction}

In the popular framework of galaxy formation and evolution (e.g., Hopkins et al.~2006,2008),
massive galaxies have been proposed to co-grow with their central supermassive black holes (SMBHs).
Intense starbursts are triggered by major gas-rich mergers \citep{barnes1992,hopkins2008,wuyts2010}
or violent disc instabilities (VDI, Dekel et al. 2009),
which also provide the fuel for the central black hole accretion. Host galaxy and SMBH
grow coevally, experiencing starburst dominated, active galaxy nucleus (AGNs)/QSO and starburst composite and AGN dominated phases,
till the AGN feedback is strong enough to expel gas and dust, making star formation and AGN activity
itself come to an end on a short timescale and finally leaving a passively evolved galaxy \citep{sanders1996,granato2004,hopkins2006,hopkins2008,alexander2012}.
%During the evolutionary sequence (Sanders 1988; ),
During the intense star formation episode, a significant amount of dust is produced, which plays an important
role in shaping the observed spectral energy distribution (SED) of a massive evolving galaxy in different phases.
Dust absorbs most of UV and optical photons and re-emits in the far-infrared (FIR) and submillimeter (submm) wavelengths.
Starburst dominated and AGN-starburst composite systems  will therefore appear to be IR luminous, just as those observed populations:
Ultra-Luminous Infrared Galaxies (ULIRGs; Sanders \& Mirabel 1996),
Submillimetre Galaxies (SMGs; Blain et al. 2002; Chapman et al. 2005; Casey et al. 2014) and
Dust-Obscured Galaxies (DOGs; Dey et al. 2008). Studying the IR luminous galaxies
at high redshift will help understanding the extreme scenarios in the early phase of the massive galaxy evolution.

Recently, \citet{eisenhardt2012} and \citet{wu2012} discovered
a new population of hyperluminous, dust-obscured galaxies using NASA's Wide-field Infrared Survey Explorer (\wise) mission
\citep{wright2010}. They selected objects by using so-called "W1W2 dropout" method.
They selected those objects which are prominent in the \textit{WISE}
12 $\mu m$ (W3) or 22 $\mu m$ (W4) bands, and faint or undetected in the 3.4 $\mu m$ (W1) and 4.6 $\mu m$ (W2) band.
These objects are rare. In total, about 1000 such objects have been identified in all sky \citep{eisenhardt2012}.
Among them, about 150 objects have spectroscopic follow-up and have been found to be mostly at high redshift,
with redshift range from 1 to 4 \citep{wu2012,tsai2015}.
%At these redshifts, the high flux densities at 22 $\mu m$ imply extremely high IR luminosities.

In order to understand the dust properties and calculate the total luminosities of these unusual
galaxies, continuum measurements at longer wavelengths are crucial.
Wu et al. (2012) observed 14 W1W2-dropout galaxies at $z>1.7$ with
the Caltech Submillimeter Observatory (CSO) SHARC-II at 350-850 $\mu m$, with nine detections,
and observed 18 with CSO Bolocam at 1.1 mm, with five detections.
Jones et al. (2014) used SCUBA-2 (Submillimetre Common-User Bolometer Array) 850 $\mu m$ band
to observe 10 dusty, luminous galaxies at $z\sim 1.7-4.6$, with six detections.
Combined \wise photometry with \textit{Herschel} PACS and SPIRE data \citep{tsai2015}, the IR
SEDs of these objects have been found to be very different from other known populations.
Their SEDs have a high mid-IR to submm luminosity ratio, which has been suggested that their IR luminosities
are dominated by emission from hot dust. Therefore, \citet{wu2012}
referred to these galaxies as hot, dust-obscured galaxies or Hot
DOGs. They are also hyperluminous: most have luminosities well over $10^{13} L_\odot$, and
some exceed $10^{14} L_\odot$, comparable to the most luminous quasars known \citep{tsai2015,assef2015a}.
The hot dust temperature and extremely high luminosity indicate that
these objects are likely heavily obscured quasars. The recent X-ray data of several Hot DOGs observed by XMM-Newton, Chandra and NuSTAR
are consistent with the scenario of them being hyperluminous, highly obscured AGNs \citep{stern2014,piconcelli2015,assef2015b}.

Besides the heavily obscured QSOs in the center of Hot DOGs, they also likely host intense star formation,
suggested by the submm/mm detections \citep{jones2014,wu2014}. Thus Hot DOGs may represent
an AGN-starburst composite system, experiencing a transit phase from a dust obscured phase to an unobscured QSO phase.
The relative contributions of AGN and starburst, which have not been well investigated in previous works, can be
analyzed based on the detailed IR SED decomposition. Different IR SED decomposition methods have been recently
carried out to analyze ULIRGs, high-$z$ radio galaxies and QSOs in the literature
\citep{mullaney2011,Han2012a,leipski2014,drouart2014,ma2015,xu2015}.

Here we construct complete mid-IR to submm/mm SEDs of a submm-detected Hot DOG sample with 
spectroscopic redshift and use a Bayesian approach to
decompose the different dust components, separating contributions from the AGN and the starburst.
In Section \ref{sect:data}, we describe the sample selection, the photometry of \textit{Herschel} observations and the compilation of mid-IR
to submm/mm SEDs. In Section \ref{sec-sed-method}, we present our Bayesian approach for IR SED decomposition. Results
and discussions are described in Section \ref{sec:results} and \ref{sec:discussions}, respectively. We summarize our
main results in Section \ref{sec:summary}. Throughout this work we assume a flat ${\rm \Lambda}$CDM cosmology with
$H_0 = 70$ km~s$^{-1}$, $\Omega_M = 0.3$, and $\Omega_\Lambda = 0.7$.

\section{Data}\label{sect:data}

\subsection{Sample}\label{subsect:sample}

\begin{table}
\centering
\caption{The sample of Hot DOGs\label{tbl-sample}}
%\begin{center}
\begin{tabular}{lccc}
\hline
\hline
Source & R.A. & Decl. & Redshift \\
Name & (J2000) & (J2000) &  \\
\hline
  W0126$-$0529  &     01:26:11.96  &     $-$05:29:09.6  & 2.937    \\
  W0134$-$2922  &     01:34:35.71  &     $-$29:22:45.4  & 3.047    \\
  W0149+2350  &     01:49:46.16  &     +23:50:14.6  & 3.228    \\
  W0220+0137  &     02:20:52.12  &     +01:37:11.6  & 3.122    \\
  W0248+2705  &     02:48:58.81  &     +27:05:29.8  & 2.210    \\
  W0410$-$0913  &     04:10:10.60  &     $-$09:13:05.2  & 3.592    \\
  W0533$-$3401  &     05:33:58.44  &     $-$34:01:34.5  & 2.904    \\
  W0615$-$5716  &     06:15:11.07  &     $-$57:16:14.6  & 3.399    \\
  W0757+5113  &     07:57:25.07  &     +51:13:19.7  & 2.277    \\
  W0859+4823  &     08:59:29.94  &     +48:23:02.3  & 3.245    \\
  W1136+4236  &     11:36:34.31  &     +42:36:02.6  & 2.390    \\
  W1248$-$2154  &     12:48:15.21  &     $-$21:54:20.4  & 3.318    \\
  W1603+2745  &     16:03:57.39  &     +27:45:53.3  & 2.633    \\
  W1814+3412  &     18:14:17.30  &     +34:12:25.0  & 2.452    \\
  W1835+4355  &     18:35:33.71  &     +43:55:49.1  & 2.298    \\
  W2054+0207  &     20:54:25.69  &     +02:07:11.0  & 2.520    \\
  W2201+0226  &     22:01:23.39  &     +02:26:21.8  & 2.877    \\
  W2210$-$3507  &     22:10:11.87  &     $-$35:07:20.0  & 2.814    \\
  W2216+0723  &     22:16:19.09  &     +07:23:54.5  & 1.680    \\
  W2238+2653  &     22:38:10.20  &     +26:53:19.8  & 2.405    \\
  W2246$-$0526  &     22:46:07.57  &     $-$05:26:35.0  & 4.593    \\
  W2305$-$0039  &     23:05:25.88  &     $-$00:39:25.7  & 3.106    \\
\hline
\end{tabular}
\end{table}

The Hot DOGs studied here are selected from the \wise All-Sky Source catalog \footnote{http://wise2.ipac.caltech.edu/docs/release/allwise/}, which provides PSF-fitting magnitudes and uncertainties in the Vega system \citep{cutri2013}. 
%The basic idea of Hot DOG selection is to search for more heavily obscured galaxies at high redshift ($z>1.5$), whose W1 (3.4 $\mu$m) and W2 (4.6 $\mu$m), sampling the rest-frame near-infrared obscuration, are faint or undetected by \wise, but whose W3 (12 $\mu$m) and W4 (22 $\mu$m) emission, tracing the hot dust heated by starbursts and/or AGNs, are well detected with SNR>5. 
The detailed selection criteria are W1 > 17.4~ (<34 $\mu$Jy) and either W4 < 7.7 (>6.9 mJy) and W2 $-$ W4 > 8.2, or W3 < 10.6 (>1.7 mJy) and W2 $-$ W3 > 5.3 \citep{eisenhardt2012,wu2012}. With several additional constraints, the resulting sample contains 934 objects over approximately 32,000 deg$^2$ \citep{assef2015a}.
%Additional constraints include that (1) they have at least seven individual \wise exposures available for W3 and W4 photometry, (2) that they are farther than $30^\circ$ from the Galactic center and $10^\circ$ from the Galactic plane in order to limit the contamination by Galactic objects, (3) they are not artifacts flagged by the \wise pipeline and not known asteroids, (4) they pass a series of visual inspections of both individual exposures and coadded images for evidence of spurious sources \citep{eisenhardt2012}. 

In order to investigate the detailed IR SEDs of Hot DOGs, we select a subsample of 22 objects (Table 1) from the full sample.
We require that all of them have known spectroscopic redshift $z>1.5$ in the literature \citep{wu2012,jones2014,tsai2015}. We also require that
they have both \textit{Herschel} PACS and SPIRE observations and have either SPIRE 500 $\mu$m or SCUBA-2 850 $\mu$m detection, 
which corresponds to >100 $\mu$m at rest frame.
%We concern ourselves with the question how significant the contribution of cold dust emission to the total IR
%energy output would be for these submm-detected Hot DOGs.
By imposing the submm detection, we will select those objects with 7 and even more detections between the observed 12 $\mu$m and millimeter bands.
Thanks to the submm detection, the properties of cold dust component can be well constrained, such as IR luminosity and cold dust 
temperature (see Table \ref{tbl-para},\ref{tbl-lum}), according to the IR SED decomposition. 
We notice that we are most likely biasing our sample towards the most intense star forming systems. We can quantify the maximal contribution of star formation in this specific class of objects and its contribution to the total IR luminosity. We are therefore insured to estimate meaningful upper limits on the expected maximal star formation contribution for Hot DOGs.

\subsection{Photometry}\label{sed-phot}

\begin{table*}
\centering
\caption{Photometry of Hot DOGs\label{tbl-phot}}
%\begin{center}
\begin{tabular}{lccccccccc}
\hline
\hline
Source & 12 $\mu$m & 22 $\mu$m & 70 $\mu$m & 160 $\mu$m & 250 $\mu$m & 350 $\mu$m & 500 $\mu$m & 850 $\mu$m & 1100 $\mu$m \\
Name & (mJy) & (mJy) & (mJy) & (mJy) & (mJy) & (mJy) & (mJy) & (mJy) & (mJy)  \\
\hline
  W0126$-$0529  &     1.0 $\pm$   0.1  &    27.2 $\pm$   1.2  &    34.8 $\pm$  3.5  &   231.8 $\pm$ 10.8  &   204.6 $\pm$  5.7  &   132.6 $\pm$  6.9  &    61.7 $\pm$  7.0  &   $...$  &   $...$  \\
  W0134$-$2922  &     4.6 $\pm$   0.2  &    19.0 $\pm$   1.4  &    24.5 $\pm$  4.6  &    41.8 $\pm$  8.9  &    49.1 $\pm$  6.1  &    26.9 $\pm$  6.5  &    36.3 $\pm$  7.7  &   $...$  &   $...$  \\
  W0149+2350  &     1.8 $\pm$   0.1  &     9.2 $\pm$   0.8  &    29.0 $\pm$  4.0  &    56.4 $\pm$ 11.0  &    42.9 $\pm$  5.4  &    38.7 $\pm$  6.9  &    33.9 $\pm$ 10.3  &   $...$  &     2.0 $\pm$  0.4 $^a$  \\
  W0220+0137  &     1.8 $\pm$   0.1  &    12.0 $\pm$   0.8  &    65.4 $\pm$  3.6  &   119.0 $\pm$ 10.8  &    95.0 $\pm$  5.0  &    77.9 $\pm$  7.0  &    39.0 $\pm$  6.4  &   $...$  &     6.2 $\pm$  2.0 $^b$  \\
  W0248+2705  &     2.0 $\pm$   0.1  &    11.1 $\pm$   1.0  &    21.5 $\pm$  4.8  &    81.6 $\pm$ 15.1  &    57.3 $\pm$  5.0  &    47.6 $\pm$  6.5  &    26.3 $\pm$  7.1  &   $...$  &    $<3.6$ $^b$ \\
  W0410$-$0913  &     2.5 $\pm$   0.1  &    12.4 $\pm$   1.0  &    36.0 $\pm$  4.6  &   107.9 $\pm$ 13.1  &   124.4 $\pm$  4.7  &   128.8 $\pm$  5.6  &    99.0 $\pm$  6.0  &    40.0 $\pm$ 14.0 $^c$ &    13.6 $\pm$  2.6 $^b$ \\
  W0533$-$3401  &     3.0 $\pm$   0.1  &    11.9 $\pm$   1.1  &    39.3 $\pm$  5.9  &    97.4 $\pm$ 14.0  &   107.5 $\pm$  4.8  &    76.3 $\pm$  7.3  &    48.9 $\pm$  4.5  &   $...$  &   $...$  \\
  W0615$-$5716  &     2.2 $\pm$   0.1  &    14.8 $\pm$   0.8  &    56.6 $\pm$  2.9  &    93.2 $\pm$  7.8  &    51.4 $\pm$  5.2  &    38.0 $\pm$  6.9  &    28.4 $\pm$  6.4  &   $...$  &   $...$  \\
  W0757+5113  &     1.5 $\pm$   0.1  &     9.3 $\pm$   0.8  &    20.1 $\pm$  3.5  &    32.9 $\pm$ 19.5  &    44.4 $\pm$  5.3  &    44.1 $\pm$  6.3  &    30.7 $\pm$  6.6  &   $...$  &    $<4.7$  $^b$  \\
  W0859+4823  &     2.2 $\pm$   0.1  &    11.8 $\pm$   0.9  &    37.5 $\pm$  3.8  &    33.8 $\pm$ 11.2  &    63.6 $\pm$  4.9  &    71.1 $\pm$  6.0  &    51.4 $\pm$  6.2  &   $...$  &     6.2 $\pm$  1.5 $^b$ \\
  W1136+4236  &     1.6 $\pm$   0.1  &     7.1 $\pm$   0.7  &   $<13.5$  &   101.7 $\pm$ 15.2  &    92.3 $\pm$  4.6  &    89.1 $\pm$  5.6  &    58.9 $\pm$  5.7  &     5.3 $\pm$  1.7 $^d$ &   $...$  \\
  W1248$-$2154  &     2.6 $\pm$   0.1  &    12.9 $\pm$   0.9  &    54.5 $\pm$  4.2  &    61.5 $\pm$  8.7  &    56.6 $\pm$  5.1  &    42.9 $\pm$  5.4  &    20.8 $\pm$  5.0  &   $...$  &   $...$  \\
  W1603+2745  &     3.2 $\pm$   0.1  &    10.1 $\pm$   1.0  &    13.4 $\pm$  3.3  &    66.4 $\pm$ 11.0  &    69.0 $\pm$  5.0  &    55.1 $\pm$  5.3  &    35.6 $\pm$  6.8  &    10.2 $\pm$  1.8 $^d$ &   $...$  \\
  W1814+3412  &     2.0 $\pm$   0.1  &    14.9 $\pm$   1.0  &    39.3 $\pm$  5.3  &    72.7 $\pm$ 12.5  &    66.5 $\pm$  4.7  &    48.2 $\pm$  4.6  &    31.4 $\pm$  6.7  &    $<3.6$ $^d$ &   $...$  \\
  W1835+4355  &     6.3 $\pm$   0.2  &    27.7 $\pm$   1.0  &    45.5 $\pm$  4.2  &   100.5 $\pm$ 12.5  &    94.0 $\pm$  5.0  &    80.8 $\pm$  5.6  &    38.6 $\pm$  5.4  &     8.0 $\pm$  1.5 $^d$ &   $...$  \\
  W2054+0207  &     4.2 $\pm$   0.2  &    11.6 $\pm$   1.0  &    15.6 $\pm$  4.8  &    73.7 $\pm$ 10.4  &    36.4 $\pm$  4.4  &    35.6 $\pm$  4.0  &    29.5 $\pm$  7.2  &    $<3.6$ $^d$ &   $...$  \\
  W2201+0226  &     4.5 $\pm$   0.2  &    17.8 $\pm$   1.5  &    23.6 $\pm$  4.4  &   134.6 $\pm$  9.5  &   156.1 $\pm$  5.7  &   136.2 $\pm$  7.8  &    76.0 $\pm$  6.1  &   $...$  &   $...$  \\
  W2210$-$3507  &     2.1 $\pm$   0.1  &    16.4 $\pm$   1.1  &    55.1 $\pm$  3.7  &   117.3 $\pm$ 14.6  &   123.9 $\pm$  5.9  &   126.4 $\pm$  5.1  &   101.5 $\pm$  6.0  &   $...$  &   $...$  \\
  W2216+0723  &     3.2 $\pm$   0.2  &    14.3 $\pm$   1.2  &    59.4 $\pm$  3.5  &   130.9 $\pm$  9.0  &    88.3 $\pm$  4.9  &    57.9 $\pm$  5.6  &   $<21.6$  &     5.5 $\pm$  1.6 $^d$  &   $...$  \\
  W2238+2653  &     2.3 $\pm$   0.1  &    17.1 $\pm$   1.0  &    62.3 $\pm$  5.4  &   141.7 $\pm$ 11.9  &   133.9 $\pm$  5.4  &    94.0 $\pm$  5.3  &    62.3 $\pm$  5.9  &   $...$  &     6.0 $\pm$  2.2 $^b$ \\
  W2246$-$0526  &     2.3 $\pm$   0.2  &    15.8 $\pm$   1.6  &    29.0 $\pm$  4.1  &   125.3 $\pm$ 11.6  &   104.0 $\pm$  3.9  &    78.6 $\pm$  5.8  &    52.4 $\pm$  5.2  &    11.4 $\pm$  2.1 $^d$ &   $...$  \\
  W2305$-$0039  &     3.2 $\pm$   0.2  &    24.4 $\pm$   1.4  &    23.7 $\pm$  2.7  &   128.4 $\pm$ 13.4  &   101.8 $\pm$  4.9  &    74.4 $\pm$  5.3  &    58.4 $\pm$  6.2  &   $...$  &   $...$  \\
\hline
\end{tabular}
\parbox{150mm} {
\textbf{Notes.} \\
$^{a}$ Flux density at 1.3 mm obtained by the SMA \citep{wu2014}. \\
$^{b}$ Flux density or (2$\sigma$) upper limits  at 1.1 mm  from CSO/Bolocam \citep{wu2012}. \\
$^{c}$ Flux density at 850 $\mu$m obtained by the CSO/SHARC-II \citep{wu2012}. \\
$^{d}$ Flux density or (2$\sigma$) upper limits at 850 $\mu$m from JCMT/SCUBA-2 \citep{jones2014}. \\
}
\end{table*}

The \wise W3 and W4 photometry for the Hot DOG sample discussed in this work is from the ALLWISE Data Release \citep{cutri2013}.
W3 and W4 flux densities and uncertainties (see Table 2) have been converted from catalog Vega magnitude by using zero points of
29.04 and 8.284 Jy, respectively \citep{wright2010}.

We also listed the FIR photometry of our 22 Hot DOGs obtained with the \textit{Herschel Space Observatory} \citep{pilbratt2010} in Table 2.
The \textit{Herschel} data (PI: P.R.M. Eisenhardt) include both PACS (Photodetector Array Camera and
Spectrometer; Poglitsch et al. 2010) observations at 70 $\mu$m and 160 $\mu$m and
SPIRE (Spectral and Photometric Imaging REceiver; Griffin et al. 2010)
observations at 250 $\mu$m, 350 $\mu$m and 500 $\mu$m. We retrieved the \textit{Herschel} data via \textit{Herschel} Science Archive (HSA)\footnote{http://www.cosmos.esa.int/web/herschel/science-archive}. Both PACS and SPIRE data were reduced using
the \textit{Herschel} Interactive Processing Environment (HIPE v12.1.0).
For PACS fluxes, we retrieved the PACS data from the HSA and reduced them with the provided PACS photometer pipeline for minimap and central point source in HIPE v12.1.0, leaving all options at their default values. After applying a mask as a combination of a central 20" radius mask and pixels at signal-to-noise$>$3 on the rest of the image, highpass filtering and MMT deglitching were applied on the masked scans. Finally, a mosaic was created with the two reduced scans. Aperture photometry was performed with an aperture radius of 14" (17") and a circle at 18" (36") and 24" (48") radius in the blue (red) channel to estimate the local background level. Uncertainties were calculated placing aperture in the image ($>48$") around the source. The final uncertainties were taken as the median absolute deviation of these apertures.  For SPIRE fluxes, we retrieved the pre-reduced data from the archive and applied the script to execute point source photometry directly on the level 2 maps (provided in the HIPE scripts). The SUSSextractor task was used and their associated uncertainties were derived with aperture photometry, assuming 22", 30" and 42" radius for the 250, 350 and 500 $\mu$m channels,  respectively. The uncertainties were calculated as the quadratic sum of the background fluctuation (assuming an annulus with an inner and an outer circles of 60" and 90" respectively) and the photon noise of the source in the previously calculated aperture.

Seven objects in our Hot DOG sample had JCMT SCUBA-2 850$\mu$m submm observations \citep{jones2014}. W0410$-$0913 had been detected at 850 $\mu$m with CSO SHARC-II in \citet{wu2012}. Six Hot DOGs had CSO Bolocam observations at 1.1 mm \citep{wu2012}. W0149+2350 had the 1.3 mm detection obtained by the SMA \citep{wu2014}. All the available submm and mm photometry had also been listed in Table 2.

\section{IR SED decomposition}\label{sec-sed-method}

The IR emission of Hot DOGs could come from the hotter AGN heated dust emission and/or colder young stellar population heated dust emission. To understand the principal physical processes responsible for the luminous IR emission of these galaxies, we need to determine relative contribution of the two components.Then, in order to decompose the IR SED of Hot DOGs to the two components, we need the model for each of them. For the AGN heated dust emission, which contributes mainly to the mid-IR emission, we have employed the CLUMPY torus model by Nenkova et al. (2002,2008a,b)\footnote{www.clumpy.org}. For the young stellar population heated dust emission, which contributes mainly to the FIR emission, we have employed a simple modified blackbody (MBB, or gray body) model. 

%\subsection{IR SED models: AGN clumpy torus and gray body model}
%In this work, we employ the BayeSED code to decompose the IR SEDs of Hot DOGs by using a new version of the CLUMPY torus model and a simple gray body model to represent the contribution of dust emission heated by young stellar population.

We use an updated version of the Bayesian SED fitting code  BayeSED (\citeauthor{Han2012a} \citeyear{Han2012a}, \citeyear{HanY2014a}) to decompose the IR SED of Hot DOGs by using a new version of the CLUMPY torus model and a simple gray body model to represent the contribution of dust emission heated by young stellar population. A detailed description of BayeSED can be found in Appendix \ref{bayesed}.

We use the newly calculated CLUMPY model database\footnote{http://www.pa.uky.edu/clumpy/models/clumpy\_models\_201410\_tvavg.hdf5/}. There are 1,247,400 models in the database, with 119 wavelengths for each SED. The torus-only model SEDs, which are stored in $flux_{tor}$, are used in this paper. Instead of the ANN method as employed in \citet{Han2012a}, we use KNN method to interpolate these model SEDs. As shown in \cite{HanY2014a}, the KNN method results in a better interpolation of SEDs, though it leads to a larger data file. The size of the original database, which is  provided as an HDF5 file, is $1.2$ GB. With the machine learning methods employed in BayeSED, it is reduced to only $180$ MB without noticeable loss of information (we have ignored the principal components with variation less than 0.01\% of the total, and have used the left 21 principal components). The CLUMPY torus model have 6 parameters: the number of clouds along a radial equatorial path $N_0$, the ratio of the outer to the inner radii of the toroidal distribution $Y=R_{\rm o}/R_{\rm d}$, the viewing angle measured from the torus polar axis $i$, the index $q$ of the radial density profile  $r^{-q}$, the width parameter characterizing the angular distribution $\sigma$, and the effective optical depth of clumps $\tau_{\rm V}$. The priors for the 6 parameters are assumed to be uniform distributions truncated to the following intervals: $N_0 = [1,15]$, $Y=[5,100]$, $i = [0,90]$, $q=[0,3]$, $\sigma = [15,70]$, $\tau_{\rm V} = [10,300]$.

Two more quantities have been defined by \citet{Nenkova2008a} to describe the dust covering of AGNs. One is the probability that a photon emitted by the AGN in direction of the given inclination angle of the torus with respect to the line of sight will escape the obscuring structure, or in other words, the probability that the object can be observed as a Type 1 AGN ($P_{type1}$). The other is the geometrical dust covering factor of the torus, $f_2$, which is also the average of the fraction of the AGN radiation absorbed by obscuring clouds. These two quantities can be set by the six free parameters of CLUMPY model. Assuming the optically thick clouds, $P_{type1}$ can be approximately written as a function of the inclination angle, $i$: 
\begin{equation}\label{equ:ptype1}
  P_{type1}=e^{-N_0e^{-\frac{\theta^2}{\sigma^2}}}
\end{equation}
where $\theta=\pi/2-i$. The geometrical dust covering factor, $f_2$, can be derived by integrating $P_{type1}$ and subtracting from 1 \citep{Nenkova2008a,mor2009}:
\begin{equation}\label{equ:f2}
  f_2=1-\int^{\pi/2}_0P_{type1}cos(\theta)d\theta
\end{equation}

The gray body model is defined as:
\begin{equation}\label{equ:gb}
  S_{\lambda}\propto(1-e^{-(\frac{\lambda_0}{\lambda})^{\beta}}) B_\lambda(T_{dust})
\end{equation}
%where
%\begin{equation}
%  \tau_{\lambda}=(\frac{\lambda_0}{\lambda})^{\beta}
%\end{equation},
where $B_\lambda$ is the Planck blackbody spectrum, $T_{\rm dust}$ is dust temperature, and we use the typical value of  $\lambda_0$ = 125$\,\mu$m.
We adopt $\beta$=1.6, which is the value typically used for high redshift QSOs \citep{beelen2006,wang2008,wang2011}.
So, the dust temperature $T_{dust}$ is the only free parameter with a uniform prior truncated to the  interval of $log(T_{dust}/K)=[1,2]$.

\section{Resutls}\label{sec:results}

\subsection{Model comparison}\label{subsec:model-comp}

\begin{figure}
\plotone{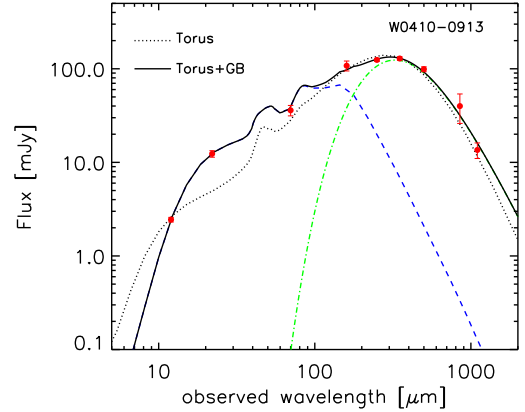}
\caption{Observed SED of a Hot DOG, W0410$-$0913 (filled circles) together with model fitting. The dotted line shows a torus-only model fit and the solid line represents the combined Torus+GB model. The dashed and dot-dashed lines represent the components of torus and gray body in Torus+GB model, respectively. As listed in Table \ref{tbl-ev}, its natural logarithm of Bayes factor \citep{jeffreys1961} ${\rm ln} (\frac{{\rm ev_{TORUS+GB}}}{{\rm ev_{TORUS}}}) =37.37$ represents a strong evidence in favor of Torus+GB model.}\label{fig:mod_comp}
\end{figure}

Previous works found that the IR SEDs of Hot DOGs are very similar, showing a steep spectrum at $1-10\mu m$ which is due to the selection criteria of Hot DOGs. Compared to various galaxy SED templates in Polletta et al. 2007, such as Arp 220 (starburst galaxy), Mrk 231 (heavily-obscured AGN and starburst composite), QSO 1 and QSO 2 (optically selected Type 1 and Type 2 QSOs), the mid-IR to submm SEDs of Hot DOGs appear to be flatter  \citep{wu2012,jones2014,tsai2015}. The obvious difference between the SEDs of Hot DOGs and the compared galaxy SED templates prompts us to fit the IR SEDs of Hot DOGs with other models.

At first, we use a torus-only model of Nenkova et al. (2002,2008a,2008b), as presented in the CLUMPY library (noted as hereafter Torus) to fit the IR SEDs of all Hot DOGs. Then we use a combined model, torus plus a gray body (Torus+GB) component, to do the SED decomposition. The presence of significant star formation activity in Hot DOGs has also been suggested by \citet{frey2016}. They found that the sum of the VLBI component flux densities is always smaller than the total flux density, indicating that star formation activity in the host galaxy should be responsible for the missing flux density. In Figure \ref{fig:mod_comp}, we show an example of IR SED fitting results with Torus (dotted line) and Torus+GB model (solid line), respectively. In the case of W0410$-$0913 (Figure \ref{fig:mod_comp}), Torus+GB model seems to provide a better fitting to the observations than Torus model. However, Torus+GB model also introduces one more free parameter than Torus model.

In order to compare different models quantitatively, we derive their Bayesian evidences, which represent a practical implementation of the Occam's razor principle. In our case, Torus+GB model having more parameters will have a lower Bayesian evidence unless it provides a significantly better fitting than Torus model. In Table \ref{tbl-ev}, we present the natural logarithm ${\rm ln(ev_{TORUS})}$ and ${\rm ln(ev_{TORUS+GB}})$ of the Bayesian evidences for Torus and Torus+GB models. We also present the natural logarithm of Bayes factor ${\rm ln} (\frac{{\rm ev_{TORUS+GB}}}{{\rm ev_{TORUS}}}$) in Table \ref{tbl-ev}. We find that the Torus+GB model has the higher Bayesian evidence than the Torus model for all Hot DOGs. We also find that ${\rm ln} (\frac{{\rm ev_{TORUS+GB}}}{{\rm ev_{TORUS}}}$) $> 10$ (corresponding to odds of $> 20000:1$), which represents strong evidence in favor of Torus+GB model according to the empirically calibrated Jeffreys's scale \citep{jeffreys1961,trotta2008}. Thereafter, we use the results of the SED fitting with Torus+GB model.

\subsection{Model parameters}

\begin{figure*}
\plotone{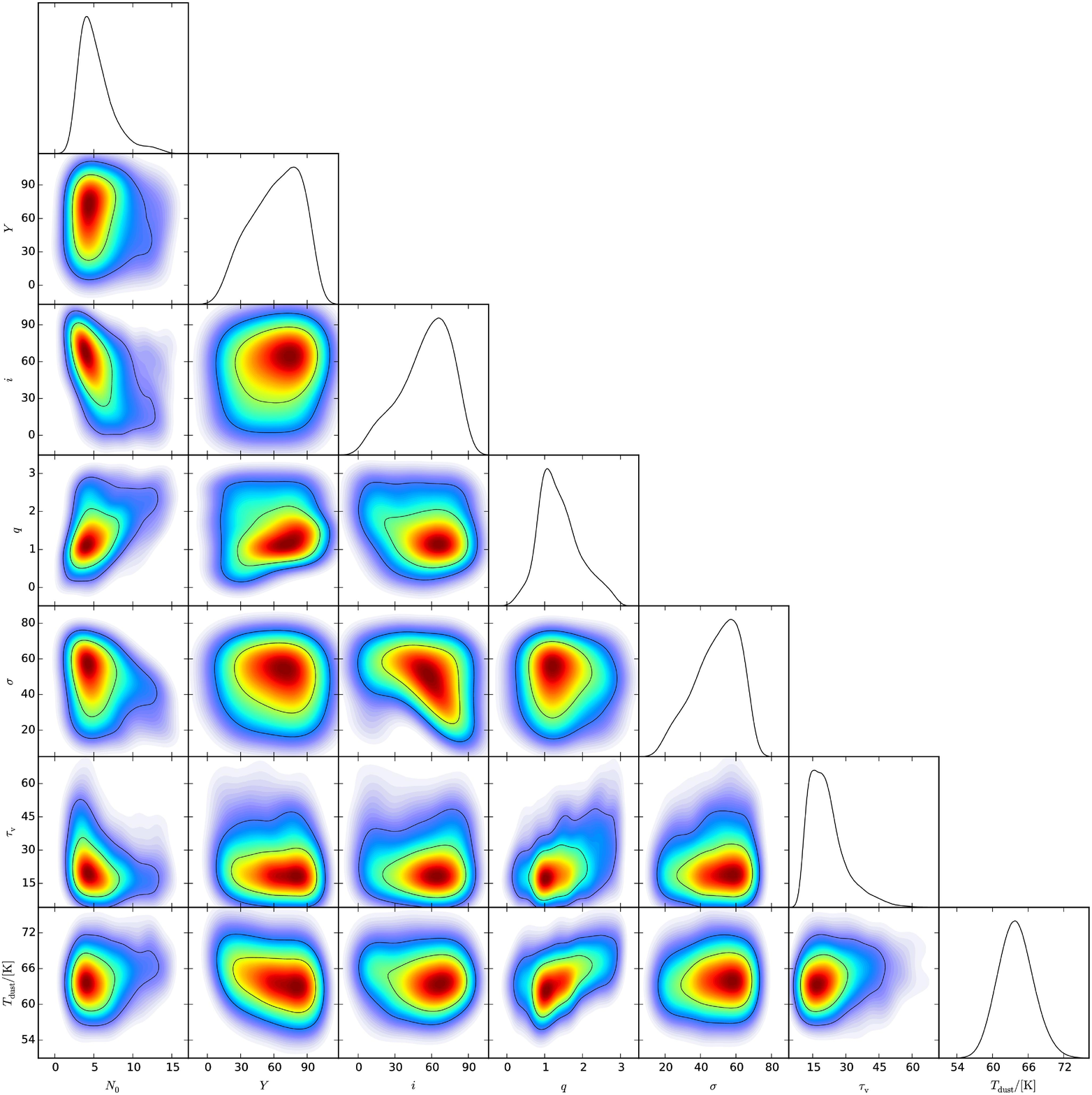}
\caption{One- and two-dimensional marginalized posterior probability distributions of the 7 free parameters,
    including 6 free parameters ($Y,i,q,\sigma,N_0,\tau_V$) for torus model and 1 free parameter ($T_{dust}$) for
    gray body model, for the Hot DOG, W0410$-$0913. The colour coding represents confidence levels.
    Both one- and two-dimensional marginalized posterior probability distributions have been normalized to unit
    area.}\label{fig:freep}
\end{figure*}

\begin{figure}
\plotone{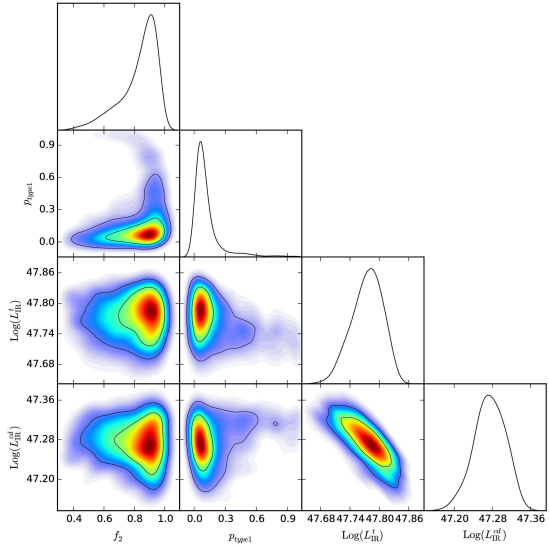}
\caption{One- and two-dimensional marginalized posterior probability distributions of four derived quantities: the geometrical
covering factor of the torus ($f_2$), the probability that light from the central source can escape the obscuring structure
without interacting with the clouds and therefore the object can be observed as a Type 1 AGN ($P_{type1}$), $1-1000\mu$m IR luminosities
of torus ($L_{IR}^{t}$) and cold dust ($L_{IR}^{cd}$) components, for W0410$-$0913 as an example.
The colour coding represents confidence levels.Both one- and two-dimensional marginalized posterior probability distributions have been normalized to unit
    area.}\label{fig:derivedp}
\end{figure}

Our Bayesian analysis of SEDs has the advantage of providing detailed posterior distribution for the free and derived parameters. From these probability distributions, we can derive the best expectations and uncertainties of all parameters. From the detailed posterior probability distributions of all parameters, it is easy to find out if a parameter is well-constrained or not. Figure \ref{fig:freep} shows the one- and two-dimensional marginalized posterior probability distributions of the 7 free parameters, including 6 free parameters ($ Y,i,q,\sigma,N_0,\tau_V$) for torus model and 1 free parameter ($T_{dust}$) for gray body model, for one Hot DOG W0410$-$0913 as an example. We can see that the gray body temperature $T_{dust}$ is tightly constrained: $T_{dust}$ is constrained to a narrow range, around $\sim63K$. However, Some parameters are loosely constrained: for example, $i$, the inclination angle of the torus with respect to the line of sight, and $Y$, the ratio between the radius of the torus and the dust sublimation radius, are rather weakly constrained.
%Both of $i$ and $\sigma$  reflect the possibility of dust obscuration and they are two degenerate parameters in general.

Figure \ref{fig:derivedp} shows the one- and two-dimensional marginalized posterior probability distributions off our derived quantities: the geometrical covering factor of the torus ($f_2$), the probability that light from the central source can escape the obscuring structure without interacting with the clouds and therefore the object can be observed as a Type 1 AGN ($P_{type1}$), $1-1000\mu$m IR luminosities of torus ($L_{IR}^{t}$) and cold dust ($L_{IR}^{cd}$) components, for W0410$-$0913 as an example. Both $L_{IR}^{t}$ and $L_{IR}^{cd}$ are well constrained to a narrow range. And $L_{IR}^{t}$ and $L_{IR}^{cd}$ are strongly anti-correlated. The nearly linear anti-correlation between $L_{IR}^{t}$ and $L_{IR}^{cd}$ indicates that $L_{IR}^{tot}$, the sum of $L_{IR}^{t}$ and $L_{IR}^{cd}$, is tightly constrained by the observed SEDs.

In order to give a good estimate for all parameters and their spreads, we use the median and percentile statistics. The lower and upper quartiles are the values below which 16\% and 84\% of points fall, respectively. We list the median values and 16\% and 84\% quartiles of seven free parameters ($Y,i,q,\sigma,N_0,\tau_V, T_{dust}$) and two derived quantities ($f_2,P_{type1}$) in Table \ref{tbl-para}. The other two derived quantities ($L_{IR}^{t}$ and $L_{IR}^{cd}$) are separately listed in Table \ref{tbl-lum} (see also Section \ref{subsec-lum-est}).

\subsection{Luminosity Estimates}\label{subsec-lum-est}

\begin{figure*}
\plottwo{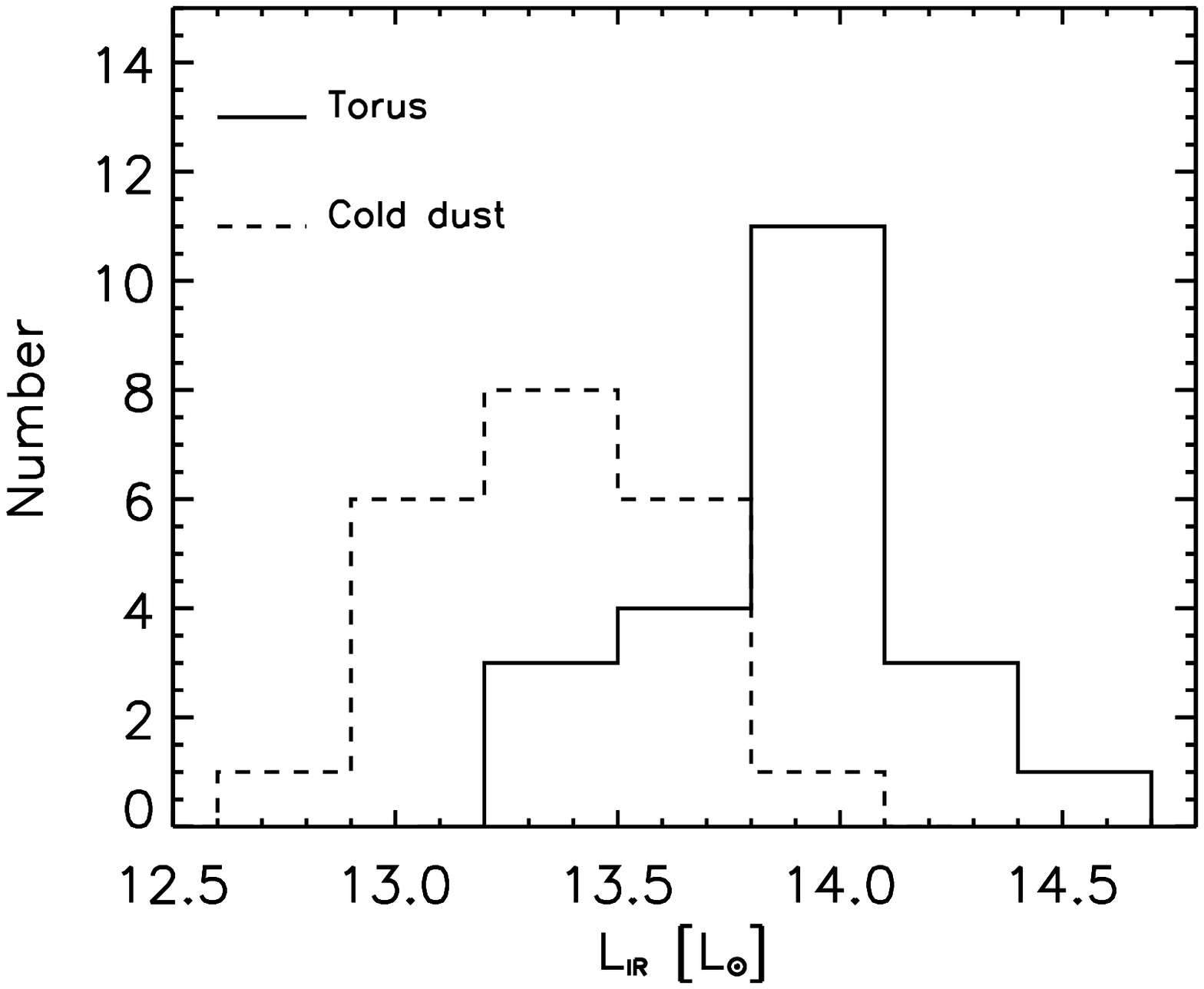}{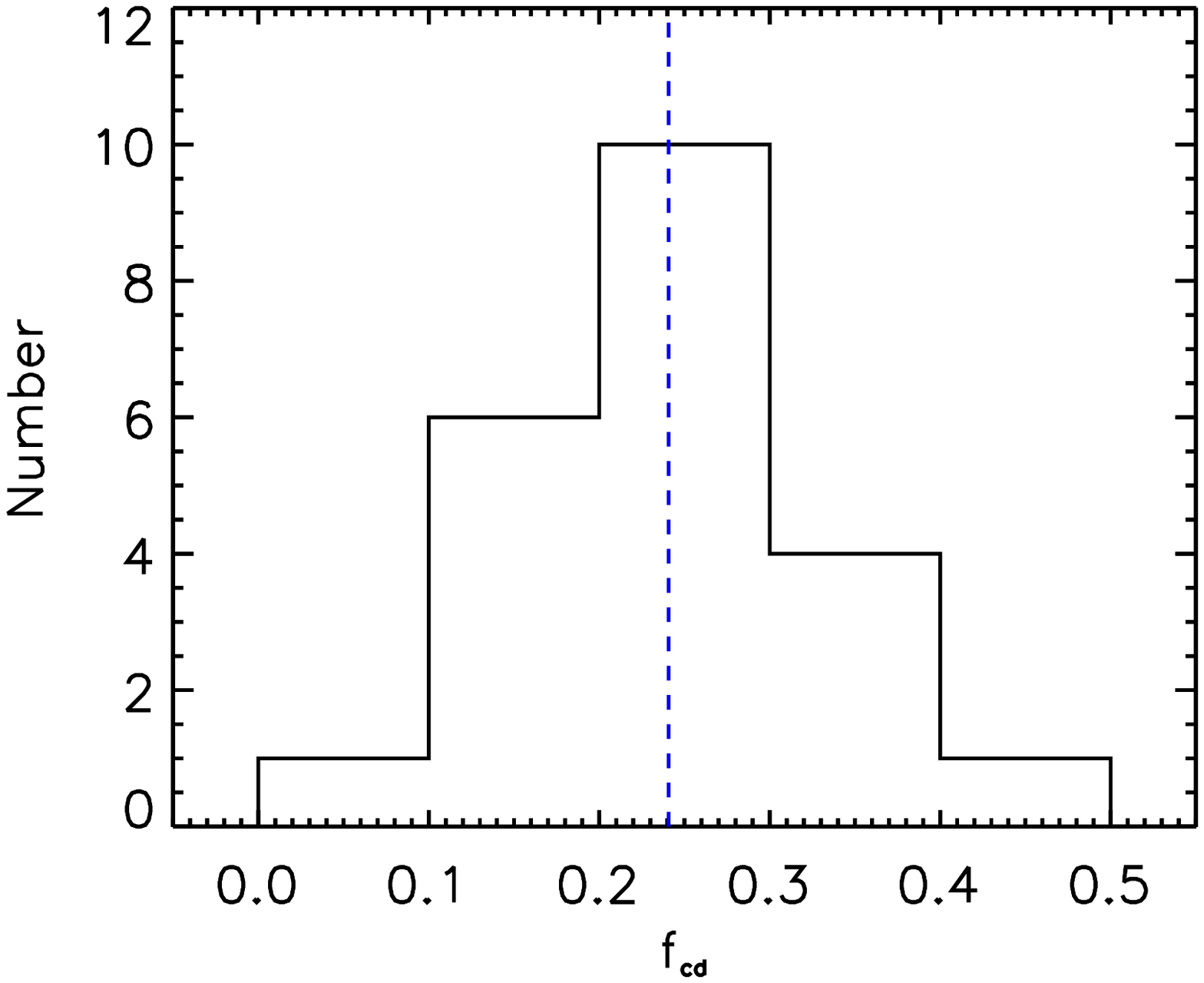}
\caption{Left: IR luminosity distributions of torus (solid line) and cold dust components (dashed line).
Right: Distribution of the fraction of cold dust component to the total IR luminosity ($f_{cd}$=$\frac{L_{IR}^{cd}}{L_{IR}^{cd}+L_{IR}^t}$). Dashed line marks the median value of $f_{cd}$ at 0.24.}\label{fig:distlir}
\end{figure*}

\begin{table}
\centering
\caption{Luminosities of Hot DOGs\label{tbl-lum}}
%\begin{center}
\begin{tabular}{lccc}
\hline
\hline
Source & log $L_{IR}^{t}$ & log $L_{IR}^{cd}$ &   log $L_{IR}^{tot}$     \\
       & ($L_\odot$) & ($L_\odot$) & ($L_\odot$) \\
\hline
  W0126-0529  &  13.98$^{+ 0.01}_{- 0.01}$  &  13.91$^{+ 0.01}_{- 0.01}$  &  14.25$^{+ 0.01}_{- 0.01}$  \\
  W0134-2922  &  14.02$^{+ 0.01}_{- 0.02}$  &  13.20$^{+ 0.04}_{- 0.04}$  &  14.08$^{+ 0.02}_{- 0.02}$  \\
  W0149+2350  &  13.89$^{+ 0.02}_{- 0.02}$  &  13.23$^{+ 0.05}_{- 0.05}$  &  13.98$^{+ 0.02}_{- 0.03}$  \\
  W0220+0137  &  14.08$^{+ 0.02}_{- 0.02}$  &  13.52$^{+ 0.06}_{- 0.07}$  &  14.19$^{+ 0.03}_{- 0.03}$  \\
  W0248+2705  &  13.45$^{+ 0.02}_{- 0.02}$  &  13.06$^{+ 0.05}_{- 0.05}$  &  13.60$^{+ 0.03}_{- 0.03}$  \\
  W0410-0913  &  14.20$^{+ 0.02}_{- 0.02}$  &  13.70$^{+ 0.03}_{- 0.02}$  &  14.31$^{+ 0.02}_{- 0.02}$  \\
  W0533-3401  &  13.88$^{+ 0.02}_{- 0.02}$  &  13.54$^{+ 0.03}_{- 0.03}$  &  14.05$^{+ 0.02}_{- 0.02}$  \\
  W0615-5716  &  14.24$^{+ 0.01}_{- 0.01}$  &  13.09$^{+ 0.13}_{- 0.14}$  &  14.27$^{+ 0.02}_{- 0.02}$  \\
  W0757+5113  &  13.42$^{+ 0.02}_{- 0.02}$  &  12.79$^{+ 0.03}_{- 0.03}$  &  13.52$^{+ 0.02}_{- 0.02}$  \\
  W0859+4823  &  14.00$^{+ 0.01}_{- 0.01}$  &  13.32$^{+ 0.01}_{- 0.02}$  &  14.08$^{+ 0.01}_{- 0.01}$  \\
  W1136+4236  &  13.61$^{+ 0.08}_{- 0.08}$  &  13.23$^{+ 0.04}_{- 0.04}$  &  13.76$^{+ 0.07}_{- 0.07}$  \\
  W1248-2154  &  14.13$^{+ 0.01}_{- 0.01}$  &  13.32$^{+ 0.05}_{- 0.04}$  &  14.19$^{+ 0.02}_{- 0.02}$  \\
  W1603+2745  &  13.61$^{+ 0.02}_{- 0.02}$  &  13.24$^{+ 0.03}_{- 0.03}$  &  13.77$^{+ 0.02}_{- 0.02}$  \\
  W1814+3412  &  13.72$^{+ 0.02}_{- 0.02}$  &  13.18$^{+ 0.04}_{- 0.04}$  &  13.83$^{+ 0.02}_{- 0.02}$  \\
  W1835+4355  &  13.89$^{+ 0.01}_{- 0.01}$  &  13.29$^{+ 0.02}_{- 0.03}$  &  13.99$^{+ 0.01}_{- 0.01}$  \\
  W2054+0207  &  13.67$^{+ 0.02}_{- 0.02}$  &  13.09$^{+ 0.05}_{- 0.05}$  &  13.77$^{+ 0.02}_{- 0.03}$  \\
  W2201+0226  &  13.92$^{+ 0.01}_{- 0.02}$  &  13.71$^{+ 0.01}_{- 0.01}$  &  14.13$^{+ 0.01}_{- 0.01}$  \\
  W2210-3507  &  13.97$^{+ 0.01}_{- 0.01}$  &  13.47$^{+ 0.01}_{- 0.01}$  &  14.09$^{+ 0.01}_{- 0.01}$  \\
  W2216+0723  &  13.37$^{+ 0.03}_{- 0.03}$  &  13.15$^{+ 0.04}_{- 0.04}$  &  13.58$^{+ 0.03}_{- 0.04}$  \\
  W2238+2653  &  13.83$^{+ 0.01}_{- 0.01}$  &  13.47$^{+ 0.02}_{- 0.02}$  &  13.99$^{+ 0.02}_{- 0.02}$  \\
  W2246-0526  &  14.46$^{+ 0.01}_{- 0.02}$  &  13.73$^{+ 0.04}_{- 0.05}$  &  14.53$^{+ 0.02}_{- 0.02}$  \\
  W2305-0039  &  14.03$^{+ 0.01}_{- 0.01}$  &  13.61$^{+ 0.02}_{- 0.02}$  &  14.17$^{+ 0.01}_{- 0.01}$  \\
\hline
\end{tabular}
\end{table}

We derive the IR luminosities of Hot DOGs based on the best-fitting results employing Torus+GB model. Our IR SED decomposition approach enables us to derive the contributions of both torus and cold dust components to the total IR energy output. In Table \ref{tbl-lum}, we listed the torus ($L_{IR}^{\rm t}$), cold dust ($L_{IR}^{cd}$) and total ($L_{IR}^{tot}$) IR luminosities within $1-1000\mu$m range. Twelve out of twenty-two Hot DOGs have $L_{IR}^{tot}>10^{14} L{_\odot}$, which are broadly consistent with the conservative estimates of IR luminosities in \citet{tsai2015}. Following \citet{tsai2015}, they are "extremely luminous infrared galaxies" (ELIRGs). The rest ten Hot DOGs have $L_{IR}^{tot}>10^{13.5} L{_\odot}$, which are hyperluminous infrared galaxies (HyLIRGs). Both the distributions of $L_{IR}^{t}$ and $L_{IR}^{cd}$ span one order of magnitude with $L_{IR}^{t}\sim10^{13.4-14.5} L_\odot$ and $L_{IR}^{cd}\sim10^{12.8-13.9} L_\odot$ and peak at $10^{13.9} L_\odot$ and $10^{13.3} L_\odot$, respectively (see Figure \ref{fig:distlir}). The torus IR luminosities of Hot DOGs are on average three times higher than those of cold dust. The fraction of cold dust component to the total IR luminosity ($f_{cd}$, see right panel in Figure \ref{fig:distlir}) ranges from 0.05 to about 0.5 with a median value of 0.24. This result confirms the previous argument that the IR energy output of Hot DOGs is dominated by hot dust emission in AGN torus. 
%However, the cold dust component is non-negligible, at least in our submm-detected Hot DOG sample.

We reminder that the relative contribution of the cold dust component is dependent on the choice of torus model. For instance, \citet{Siebenmorgen2015} presented a self-consistent AGN torus model (thereafter S15 model)\footnote{www.eso.org/$\sim$rsiebenm/agn$\_$models/} with a different chemical dust composition and grain geometries, predicting that the AGN torus would have stronger FIR/submm emission than that of the CLUMPY torus model. In this case, the contribution of the cold dust component will be lower than what we have derived. In order to test the effect of different models on the derived cold dust contribution, we replace the CLUMPY torus model with the S15 model and do the fitting again. We find that the Torus+GB model always has the highest Bayesian evidence among the sole S15, S15+GB and Torus+GB models. As expected, the median value of $f_{cd}$ derived from the S15+GB model is much lower ($\sim0.1$). Thus we adopt the results of the Torus+GB model and treat our estimation of $f_{cd}$ with the Torus+GB model as maximum possible value.

\subsection{Median SED of Hot DOGs}\label{subsec:composite_sed}

\begin{figure}
\plotone{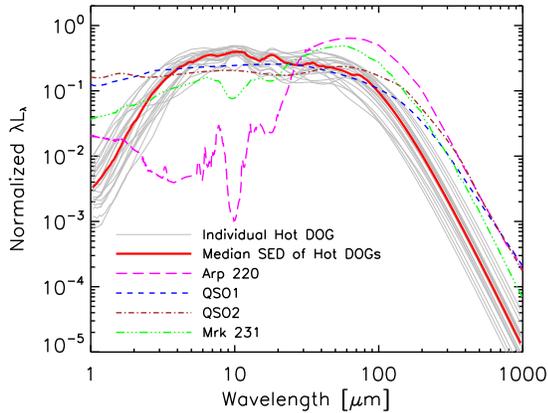}
\caption{Normalized rest-frame SEDs (gray thin lines) and the median SED (red thick line) of the submm-detected Hot DOGs.
The SEDs are based on the best-fitting with Torus+GB model and normalized to the total IR luminosity $L^{\rm tot}_{IR}$. The median
SED is derived by taking the median value of all 22 best-fitting SEDs. Individual SEDs and the median SED
have been compared to other templates, including Type 1 QSOs (QSO1), Type 2 QSOs (QSO2), a starburst galaxy, Arp 220 and
a heavily obscured AGN-starburst composite, Mrk 231 from \citet{polletta2007}. }\label{fig:SED_compare}
\end{figure}

In Figure \ref{fig:SED_compare}, we plot the rest-frame SEDs of 22 Hot DOGs based on the best-fitting with Torus+GB model.
The rest-frame SEDs have been normalized to the total IR luminosity $L^{\rm tot}_{IR}$. Then we derive a median SED by taking
the median value of 22 normalized rest-frame SEDs of Hot DOGs. The median SED of the submm-detected Hot DOGs shows
several features consistent with previous works \citep{wu2012,jones2014,tsai2015}. It has a very steep $1-5\mu$m spectrum, which
could be due to the selection effect of Hot DOGs. It becomes rather flat within the wavelength range of $\sim10-50\mu$m
where the torus emission dominates the energy output. Then it sharply drops at $>100\mu$m.

We also compare the median SED of the submm-detected Hot DOGs
with other known templates from \citet{polletta2007}, including Type 1 QSOs (QSO1), Type 2 QSOs (QSO2), a starburst galaxy, Arp 220 and
a heavily obscured AGN-starburst composite, Mrk 231. We can find that Hot DOGs have the highest luminosity ratio
between mid-IR and submm at rest-frame compared to other templates. The relatively weak emission at $>100\mu$m in the Hot DOG
median SED may be due to them having higher temperature of cold dust, which will also be suggested in Section \ref{subsec:tdust-lir}.
Within the wavelength range of $\sim6-50\mu$m, the median SED
of Hot DOGs is very similar to those of QSO1 and QSO2. This result supports the argument that Hot DOGs
are the heavily dust-obscured QSOs.

\section{Discussions}\label{sec:discussions}

\subsection{The $T_{dust}-L_{IR}$ relation}\label{subsec:tdust-lir}

\begin{figure*}
\plotone{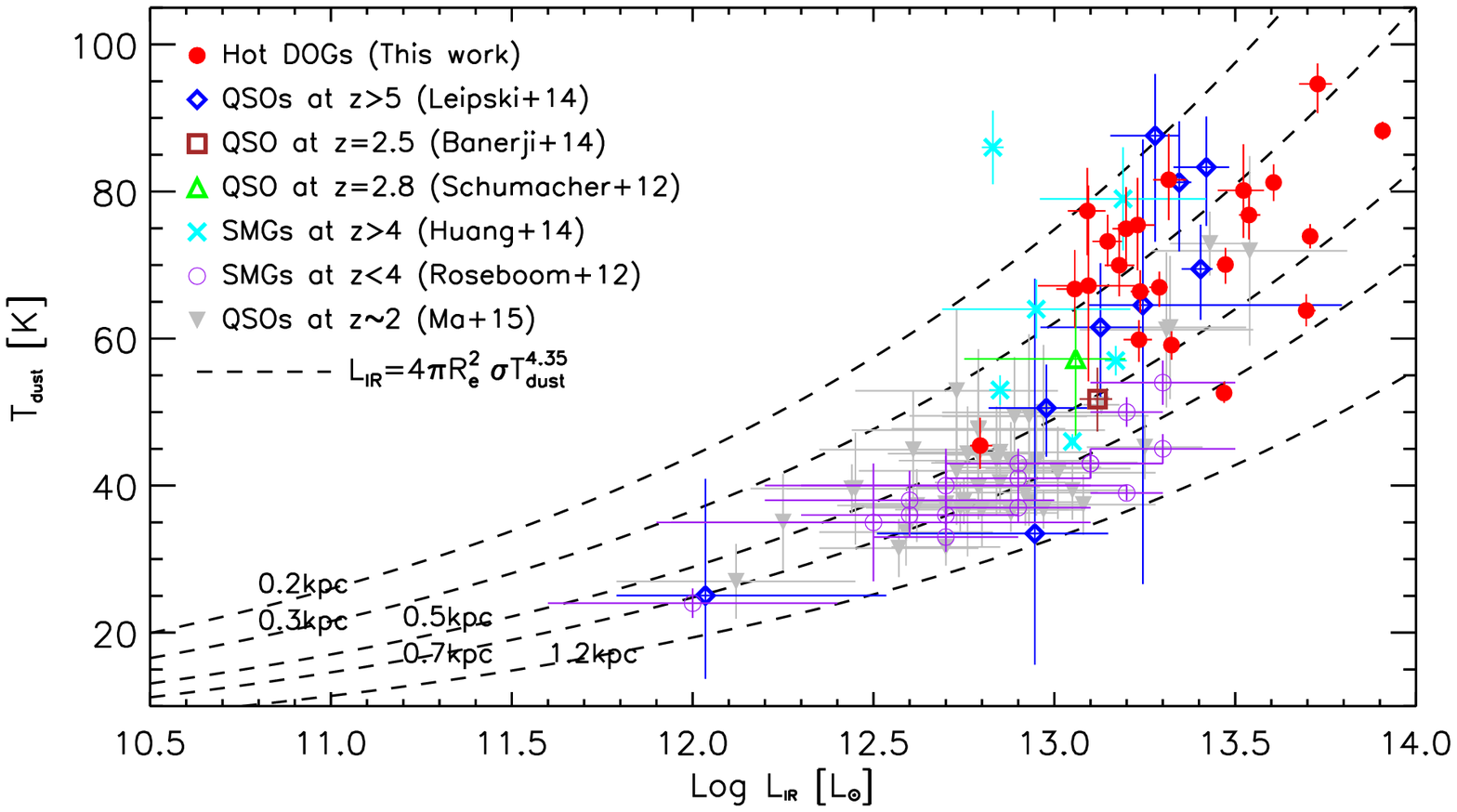}
\caption{Cold dust temperature as a function of IR luminosity for our Hot DOG sample and other high redshift
  populations: SMGs at $z<4$ \citep{roseboom2012} and $z>4$ \citep{huang2014},
  QSOs at $z>1.5$ \citep{ma2015} and $z>5$ \citep{leipski2013,leipski2014}, a very red Type 1 QSO ULASJ1234+0907 at $z=2.5$ \citep{banerji2014} and a heavily-obscured QSO AMS12 at $z=2.8$ \citep{schumacher2012}.
  The dashed lines represents the expected $T_{dust}-L_{IR}$ relation by Equation \ref{equ:mbb} with
  several different $R_e$ values (0.2, 0.3, 0.5, 0.7 and 1.2 kpc).}\label{fig:Td2LFIR}
\end{figure*}

The cold dust temperature $T_{dust}$ for the Hot DOG sample has been derived as described in Section \ref{sec-sed-method} (see also Table \ref{tbl-para}).
We note that the formula of gray body in Equation \ref{equ:gb} is for general opacity. Adopting general opacity, the dust temperatures of Hot DOGs range from 45 to 95K with a median value of about 72K. For some previous studies on SMGs (e.g., Yang et al. 2007; Lapi et al. 2011), the optically thin regime has been assumed and the term $(1-e^{-(\frac{\lambda_0}{\lambda})^\beta})$ in Equation \ref{equ:gb} can be simplified as $(\frac{\lambda_0}{\lambda})^\beta$ at $\lambda\gg \lambda_0$. The different assumption on the optical depth results in the differences of the derived dust temperatures. The dust temperatures derived with general opacity are higher than those with the optical thin assumption \citep{conley2011,magids2012,huang2014}. We test how the choice of opacity will affect the estimation of the cold dust temperature. Assuming optically thin case and using $S_{\lambda}\propto {\lambda}^{-\beta} B_\lambda(T_{dust})$ to describe GB component, we derive the dust temperatures of Hot DOGs ranging from 35 to 74K with a median value of about 49K, which is on average 23K lower than that of general opacity. Under the optically thin assumption, the derived dust temperatures remain averagely hotter than those found in ULIRGs, SMGs and DOGs which mostly range from 20K to 50K \citep{kovacs2006,magids2010,magnelli2012,melbourne2012}. 

In Figure \ref{fig:Td2LFIR}, we plot the relation between the cold dust temperature $T_{dust}$ and the IR luminosity of cold dust. We compare our sample with other populations: SMGs at $z<4$ \citep{roseboom2012} and $z>4$ \citep{huang2014}, QSOs at $z>1.5$ \citep{ma2015} and $z>5$ \citep{leipski2013,leipski2014}, a very red Type 1 QSO ULASJ1234+0907 at $z=2.5$ \citep{banerji2014} and a heavily-obscured QSO AMS12 at $z=2.8$ \citep{schumacher2012}. As all the compared samples used the gray body with the general opacity form in Equation \ref{equ:gb}, the comparison between them in Figure \ref{fig:Td2LFIR} should be self-consistent. The adopted parameters of $\beta$ and $\lambda_0$ are slightly different for each sample. $\beta$=1.8 and $\lambda_0=100~\mu$m had been used in SMGs at $z<4$ \citep{roseboom2012} while \citet{huang2014} used $\beta$=2.0 and $\lambda_0=100~\mu$m for their SMG sample at $z>4$. We selected those $500~\mu$m-detected ($\sigma>3$) QSOs with $z>1.5$ from  $250~\mu$m-detected ($\sigma>5$)
optical-selected QSO sample in \citet{ma2015}. They adopted $\beta$=1.5, which is same as the default value in \citet{casey2012}, and $\lambda_0=100~\mu$m following \citet{draine2006}. For QSOs at $z>5$, we selected nine QSOs with 500 $\mu$m and/or 1.2mm detected from \citet{leipski2013,leipski2014}. For nine $z>5$ submm/mm-detected QSOs, the very red Type 1 QSO ULASJ1234+0907 and the heavily-obscured QSO AMS12, we re-fitted their IR SEDs with Torus+GB model, adopting $\beta$=1.6 and $\lambda_0=125~\mu$m as we did for our Hot DOG sample.

The locus on the $T_{dust}-L_{IR}$ diagram of our Hot DOG sample is consistent with that of submm-detected QSOs in the similar IR luminosity range ($L_{IR}>10^{13}L_\odot$). However, compared to SMGs in the similar redshift range \citep{roseboom2012}, our submm-detected Hot DOGs are more luminous (the median value of $log L_{IR} [L_\odot]$: 13.3 vs. 12.9) and have hotter dust temperature (the median value of $T_{dust}$: 72K vs. 40K). Both the red Type 1 QSO ULASJ1234+0907 and the heavily-obscured QSO AMS12 seem to follow the $T_{dust}-L_{IR}$ relation of our submm-detected Hot DOGs, while they have slightly low temperature and IR luminosities. Interestingly, submm-detected QSOs have the same $T_{dust}-L_{IR}$ relation as SMGs in the similar redshift range at $L_{IR}\leq10^{13}L_\odot$, which indicates that they may have the similar dust properties.

In order to understand the $T_{dust}-L_{IR}$ relation of our submm-detected Hot DOGs and other populations, we try to interpret the observed $T_{dust}-L_{IR}$ relation using Stefan-Boltzmann law following \citet{symeonidis2013} and \citet{ma2015}. We note that the Stefan-Boltzmann law has the form of $L=4\pi R^2\sigma T^4$ for a perfect blackbody. While we adopt a gray body in this work, we expect that the $T_{dust}-L_{IR}$ relation will have a different form against the perfect blackbody. Following \citet{ma2015}, we integrate Equation \ref{equ:gb} and find that the $T_{dust}-L_{IR}$ relation can be described approximately by the form: 
\begin{equation}\label{equ:mbb}
L_{IR}=4\pi R_e^2\sigma T^\alpha
\end{equation}
$R_e$ can be treated as the effective radius of the equivalent FIR-emitting region.
We also find that the index $\alpha$ is dependent on the choice of the dust temperature range.
For low dust temperature ($T_{dust}$<35K), the index $\alpha$ equals 5.05, while the value decreases to
4.35 for $T_{dust}\geq$ 35K. The value 4.35 of the index $\alpha$ is very close to the adopted value 4.32 in \citet{ma2015}.
The slight difference of the derived $\alpha$ can rise from the different choices of
$\beta$ and $\lambda_0$ in Equation \ref{equ:gb} between us.
As all of our Hot DOGs and most of other populations plotted in Figure \ref{fig:Td2LFIR}
have the dust temperature greater than 35K, we therefore adopt the value 4.35 of the index $\alpha$.
We plot the $T_{dust}-L_{IR}$ relation expected by Equation \ref{equ:mbb} with several
different $R_e$ (0.2, 0.3, 0.5, 0.7 and 1.2 kpc, see dashed lines in Figure \ref{fig:Td2LFIR}).
For Hot DOGs and all other populations plotted in Figure \ref{fig:Td2LFIR} having $L_{IR}>10^{12}L_\odot$, the increase in
IR luminosity is mostly due to the increase of the dust temperature.
For instance, the $T_{dust}-L_{IR}$ relation of SMGs at $z<4$ can be
described well by Equation \ref{equ:mbb}, adopting $R_e=0.7$kpc.
Compared to SMGs at $z<4$, our Hot DOGs show higher dust temperature, but smaller $R_e$ which range from 0.2 to 0.5kpc.
Thus the increase in IR luminosity of our Hot DOGs relative to that of SMGs at $z<4$ should be
dominated by the increase in dust temperature rather than $R_e$. The increase of dust temperature
could be due to the more intense radiation field caused by more intense starburst activity and/or
buried AGN activity.

\subsection{Dust mass and Gas mass}\label{subsec:mdust}

\begin{figure}
\plotone{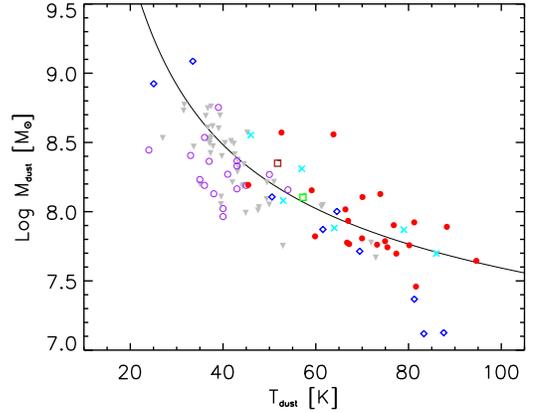}
\caption{Dust mass ($M_{dust}$) as a function of dust temperature ($T_{dust}$) for our Hot DOG sample and other high redshift
  populations. The symbols are the same as in Figure \ref{fig:Td2LFIR}. The solid line represents the $M_{dust}-T_{dust}$ relation
  at $z=3.0$ expected by Equation \ref{equ:mdust}, assuming $S_{\nu_{obs}}$=39mJy at $\nu_{obs}$=600GHz. }\label{fig:Td2mdust}
\end{figure}

Our SED fitting with Torus+GB model decomposes IR emission of Hot DOGs into hot torus and cold dust components.
The cold dust temperature has been constrained well. We can therefore estimate the mass of cold dust
using:
\begin{equation}\label{equ:mdust}
    M_{dust}=\frac{D^2_L}{(1+z)}\times\frac{S_{\nu_{obs}}}{\kappa_{\nu_{rest}} B(\nu_{rest},T_{dust})}
\end{equation}
where $D_L$ is the luminosity distance, $S_{\nu_{obs}}$ is the flux density at observed frequency $\nu_{obs}$, $\kappa_{\nu_{rest}}=\kappa_0(\nu/\nu_0)^\beta$
is the dust mass absorption coefficient at the rest frequency of the observed band, $B(\nu_{rest},T_{dust})$ is the Planck function
at temperature $T_{dust}$.  
The main uncertainty of dust mass estimation arises from the choice of the $\kappa_{\nu_{rest}}$ value. In the literature, the $\kappa_{\nu_{rest}}$ value can vary by over one order of magnitude at given frequency/wavelength: from a very high value of $\kappa_{850\mu m}$ (i.e., $\kappa_{350{\rm GHz}}$) $\sim11$ cm$^2$ g$^{-1}$ suggested by laboratory measurements and theoretical modelling, $\kappa_{850\mu m}\sim1.6-8$ cm$^2$ g$^{-1}$ from the observations of newly formed dust, to a very low value of $\kappa_{850\mu m}$ ($\sim0.4$ cm$^2$ g$^{-1}$) supported by studies of extragalactic systems and diffuse ISM dust in the Galaxy \citep{james2002,dunne2003,draine2003,Siebenmorgen2014}.
In this paper, we adopt a moderate value of $\kappa_{\rm 1THz}=20$ cm$^2$ g$^{-1}$, which is the same as in \citet{wu2014}. Given $\beta=1.6$ and $\kappa_{\rm 1THz}=20$ cm$^2$ g$^{-1}$, we can derive $\kappa_{850\mu m}=3.8$ cm$^2$ g$^{-1}$. We use the flux density at 500 $\mu$m (or 850 $\mu$m, if detected) for dust mass estimation of Hot DOGs. For high redshift SMGs and QSOs, the detected, longest-wavelength band (normally among 500, 850 or 1200 $\mu$m) has been used to estimate their dust mass.

We plot the dust mass as a function of dust temperature for our Hot DOG sample and high redshift SMGs and QSOs in
Figure \ref{fig:Td2mdust}. As we estimate the dust mass of Hot DOGs and all other populations
plotted in Figure \ref{fig:Td2mdust} adopting the same value of $\kappa_{850\mu m}$, the dust mass comparison among
them will be self-consistent. The logarithm values of dust mass (Log M$_{dust}$ [M$_\odot$]) range from
7.5 to 8.6 with a median value of 7.9 for our Hot DOG sample. The median values of Log M$_{dust}$ of both SMGs at $z<4$
(purple open circles) and QSOs at $z>1.5$ (gray triangles) are about 0.4-0.5 dex higher than those of Hot DOGs.
Our result is inconsistent with that of \citet{wu2014}. They reported that the cold dust masses of Hot DOGs are
comparable to those in submm-detected QSOs  with a median value of about $10^{8.5}$ M$_\odot$,
and a bit higher than those in SMGs. They derived the cold dust masses by assuming a fixed and lower dust
temperature ($T_{dust}=35$K). We find that the cold dust masses decrease by a significant factor
as the derived dust temperature increases by a factor of about two. In Figure \ref{fig:Td2mdust}, we also plot the
$M_{dust}-T_{dust}$ relation at $z=3.0$ expected by Equation \ref{equ:mdust}, assuming $S_{\nu_{obs}}$=39mJy at $\nu_{obs}$=600GHz.
For the $M_{dust}-T_{dust}$ relation at $z=3.0$, $M_{dust}\propto T_{dust}^{-2.3}$ at $T_{dust}\geq35$K, while $M_{dust}\propto T_{dust}^{-6.6}$ at $T_{dust}<35$K. The calculation of $M_{dust}$ can be strongly affected by $T_{dust}$.
As a result, our Hot DOGs with hotter dust temperature have lower dust masses compared to SMGs and submm-detected QSOs,
even though they have hyperluminous cold dust emissions ($L_{IR}>\sim10^{13}L_\odot$).

Molecular gas masses in Hot DOGs can be calculated from dust masses assuming a fiducial dust-to-gas ratio of Milky Way
$\sim 0.01$. The median value of molecular gas masses in Hot DOGs is about $10^{10}$M$_\odot$. As a comparison,
SMGs are more gas rich than Hot DOGs. Molecular gas masses in SMGs %and submm-detected QSOs
are about $10^{10.5}$M$_\odot$ which are consistent with the estimations by converting CO J$=1-0$ line luminosity to
molecular gas masses with a fiducial CO-to-H$_2$ factor (Bolatto et al. 2013; Carilli \& Walter 2013, see also Figure 3 in Wu et al. 2014).
The molecular gas mass in a Hot DOG, W0149+2350, is expected to be $\sim 5.5\times10^9$M$_\odot$, which
is consistent with the non-detection of CO J$=4-3$ line by CARMA in \citet{wu2014}. \citet{wu2014} gave a $2\sigma$ upper limit
on molecular gas mass for W0149+2350: ${\rm M_{H_2}<3.3\times10^{10}M_\odot}$.

\subsection{The dust covering factor}\label{subsec:covering}

\begin{figure}
\plotone{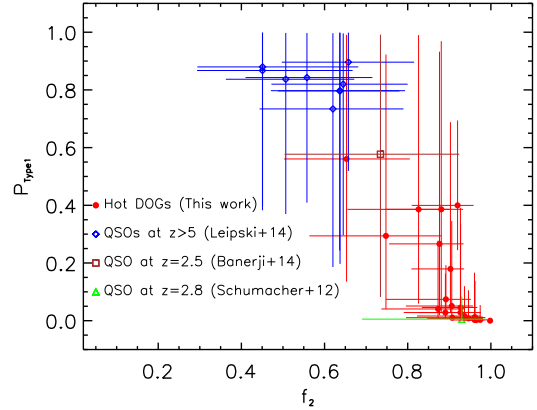}
\caption{The probability that the object can be observed as a Type 1 AGN ($P_{type1}$) as a function
of the geometrical covering factor of the torus ($f_2$), which is the ratio between the total torus luminosity and
bolometric luminosity $L_{bol}$.}\label{fig:f2_Ptype1}
\end{figure}

\begin{figure}
\plotone{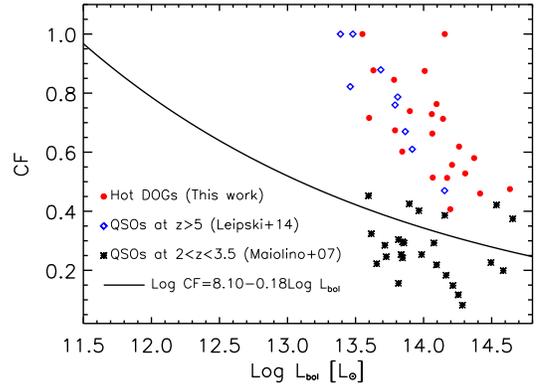}
\caption{The dust covering factor (CF) as a function of bolometric luminosity $L_{bol}$. The solid line represents the $CF-L_{bol}$ relation
derived from \citet{maiolino2007}.}\label{fig:cf}
\end{figure}

In Figure \ref{fig:f2_Ptype1}, we plot the relation between the probability that the object can be observed as a Type 1 AGN
($P_{type1}$) and the geometrical covering factor of the torus ($f_2$) of Hot DOGs and QSOs.
We emphasize that both $P_{type1}$ and $f_2$ of Hot DOGs and QSOs have been derived from the same SED fitting method with Torus+GB model,
which have been listed in Table \ref{tbl-para}. As expected by Equation \ref{equ:f2}, a clear anti-correlation
between $P_{type1}$ and $f_2$ has been seen in Figure \ref{fig:f2_Ptype1}. As a result of pre-selection,
submm-detected QSOs at $z>5$ (blue diamonds), ULASJ1234+0907 (brown square) and AMS12 (green triangle)
have been known as Type 1 QSOs, Type 1 QSO with very red color and  a heavily dust-obscured QSO, respectively.
Despite the large uncertainties, the derived values of $P_{type1}$ and $f_2$ are broadly consistent with the known inputs:
Type 1 QSOs at $z>5$ having a large value of $P_{type1}$ ($\sim 0.8-0.9$) and a moderate $f_2$ value, ULASJ1234+0907
having moderate values of both $P_{type1}$ and $f_2$ ($\sim 0.6-0.7$) and AMS12 having $P_{type1}\approx0$ and $f_2\approx1$.
The consistency indicates that our SED fitting method is able to recover dust obscuring \emph{only} based on IR SED.
All but one Hot DOGs have $P_{type1}<0.4$ and over $2/3$ Hot DOGs have $P_{type1}<0.1$.
All but two Hot DOGs have $f_2>0.8$. The low $P_{type1}$ value and high $f_2$ value confirms again that
Hot DOGs are heavily dust-obscured QSOs.

As mentioned by \citet{mor2009}, the geometrical covering factor of the torus ($f_2$), which is
the ratio between the total torus luminosity and bolometric luminosity, is different from
the apparent covering factor of the torus, which is the ratio between the observed luminosity at a given angle and
wavelength range and $L_{bol}$. The apparent covering factor can be written as
\begin{equation}\label{equ:cf}
  f(i)=\frac{1}{L_{bol}}\int^{100\mu m}_{2\mu m}L_{\lambda}d\lambda
\end{equation}
where $L_{\lambda}$ is the rest-frame monochromatic luminosity of the torus.
This definition of $f(i)$ is consistent with that of dust covering factor (CF) defined by \citet{maiolino2007}, where
CF is the ratio of thermal infrared emission to the primary AGN radiation.
We estimate the CF values of Hot DOGs by adopting $L_{bol}=BC\times L^t_{IR}$ in Equation \ref{equ:cf},
where $BC$ is a bolometric correction factor. Here we adopt $BC=1.4$ as the observed SEDs of Hot DOGs are dominated
by IR emission of torus. It is also broadly consistent with the conservative estimations of $L_{bol}$ in \citet{tsai2015}.
The CF values of submm-detected QSOs at $z>5$ have been computed by converting the mid-IR-to-optical luminosity ratio
(Equation 2 in Maiolino et al. 2007). We plot the relation between the CF values of Hot DOGs, submm-detected QSOs at $z>5$ and
$2<z<3.5$ QSOs and their bolometric luminosities in Figure \ref{fig:cf}.
The solid line represents the CF$-L_{bol}$ relation derived from \citet{maiolino2007} by combining their Equation 1 and 2.
In the literature, whether CF evolves with redshift remains controversial (e.g., Treister \& urry 2006; Hasinger 2008; Lusso
et al. 2013). However, the trend that CF decreases with increasing bolometric luminosity has been widely found locally and
at high redshift (e.g., Treister et al. 2008; Hasinger 2008; Lusso et al. 2013; Ma \& Wang 2013).
Our Hot DOGs are similar to submm-detected QSOs at $z>5$ in \citet{leipski2014}, showing a systematic offset from
the CF$-L_{bol}$ relation with respect to $2<z<3.5$ QSOs in \citet{maiolino2007}. The extremely
luminous Hot DOGs ($L_{bol}>10^{13.5}L_\odot$) have rather large dust covering factors (CF$\sim0.4-1.0$),
while $2<z<3.5$ QSOs have similar $L_{bol}$ but much lower values of CF$\sim0-0.5$.
Thus CF could span a full range of $0-1$ at $L_{bol}>10^{13.5}L_\odot$.
This result may suggest that the previously found anti-correlation between CF and $L_{bol}$ could be due to
the rare number density of found heavily-obscured QSOs at high redshift and the selection bias which may miss most heavily-obscured
QSOs in UV/optical and X-ray bands. The recent study on the most luminous AGNs at $z\sim2-3.5$ by \citet{netzer2015}
found no evidence for a luminosity dependence of the torus covering factor, which is consistent with our result.

\subsection{The coeval growth of the SMBHs and their hosts}\label{subsec:coevolution}

\begin{figure}
\plotone{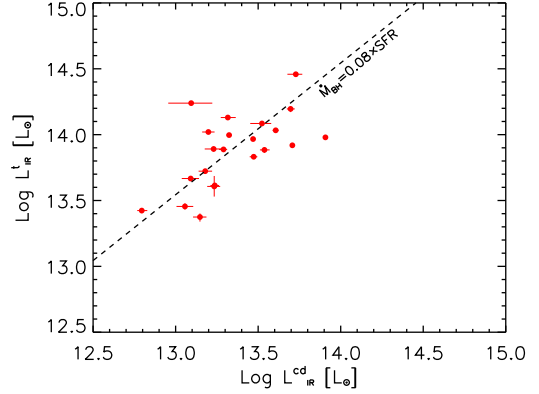}
\caption{The IR luminosity of the cold dust component $L^{cd}_{IR}$ versus the IR luminosity of torus component
$L^t_{IR}$ for Hot DOGs. The dashed line represents the observed $L^{cd}_{IR}-L^t_{IR}$ relation of Hot DOGs, which
corresponds to the relation between the star formation rate and the black hole growth rate, $\dot{M}_{BH}=0.08\times {\rm SFR}$
(see more details in Section \ref{subsec:coevolution}).}\label{fig:lc2lt}
\end{figure}

Based on the SED decompositions presented in Section \ref{sec-sed-method}, the total IR luminosities of
Hot DOGs have been disentangled into the torus and cold dust components.
Assuming that the torus and cold dust emissions are related to SMBH accretion and
star forming, respectively, the derived $L^{cd}_{IR}-L^t_{IR}$ relation as seen in Figure \ref{fig:lc2lt}
can be used to investigate the relation between SMBH accretion and star formation. In the following paragraphs,
we will describe how we convert $L^{cd}_{IR}$ and $L^t_{IR}$ into star formation rates (SFRs) and SMBH growth rate ($\dot{M}_{BH}$),
respectively.

We use the simple relation between the SFR and IR luminosity given for local galaxies \citep{kennicutt1998},
adopting a Chabrier initial mass function \citep{chabrier2003}:
\begin{equation}\label{equ:lir2sfr}
  \frac{{\rm SFR}}{M_\odot~ yr^{-1}} = 1.0\times 10^{-10}~ \frac{L^{cd}_{IR}}{L_\odot}
\end{equation}
The SFRs of Hot DOGs in our sample span from 600 to $\sim6000~ M_\odot~ yr^{-1}$, with a
median value of $\sim2000~ M_\odot~ yr^{-1}$. If adopting a Salpeter IMF \citep{salpeter1955}, the derived SFRs will increase
by a factor of 1.72. The SFRs of Hot DOGs are very high, but not rare at high redshift.
Other high-$z$ populations have the similar SFRs, such as SMGs (e.g., Chapman et al. 2005; Wardlow et al. 2011;
Casey et al. 2013; Swinbank et al. 2014; Barger et al. 2014), high-$z$ radio galaxies (e.g., Seymours et al. 2008; Barthel et al. 2012;
Rawlings et al. 2013; Drouart et al. 2014) and high-$z$ QSOs (e.g., Wang et al. 2008,2011; Leipski et al. 2013,2014; Ma \& Yan 2015).

The QSO bolometric luminosity $L_{bol}$ can be determined by black hole mass growth rate $\dot{M}_{BH}$ and radiative efficiency $\eta$,
and can also be estimated from the observed IR luminosity $L^t_{IR}$ adopting a bolometric correction factor $BC$:
\begin{equation}\label{equ:lbol_mbh}
  L_{bol} = \frac{\eta\dot{M}_{BH}c^2}{(1-\eta)} = BC\times L^t_{IR}
\end{equation}
The radiative efficiency $\eta$ varies from 0.052 for a non-rotating black hole to 0.3
for a fast rotating black hole (e.g., Shapiro 2005). We adopt the more commonly adopted value
$\frac{\eta}{1-\eta}=0.1$ (e.g., Yu \& Tremaine 2002; Marconi et al. 2004; Cao \& Li 2008).
The bolometric correction factor can vary from 1.4 to 15 for QSOs in the IR band (e.g., Elvis et al. 1994;
Marconi et al. 2004; Richards et al. 2006; Hao et al. 2014; Scott \& Stewart 2014). Here we adopt
$BC=1.4$ as the observed SEDs of Hot DOGs are dominated
by IR emission of torus. It is also broadly consistent with the conservative estimations of $L_{bol}$ in \citet{tsai2015}.

Combining Equation \ref{equ:lir2sfr} and \ref{equ:lbol_mbh} , we can convert the observed $L^{cd}_{IR}-L^t_{IR}$ relation
into the SFR$-\dot{M}_{BH}$. We derive that $\dot{M}_{BH}=0.08\times $SFR (See dashed line in Figure \ref{fig:lc2lt}).
%Considering the local spheroid mass-black hole mass relation $M_{BH}\approx0.002\times M_{sph}$,
%the black holes in Hot DOGs seem to outgrow their host galaxies by a factor of $\sim40$.

\begin{figure*}
\plotone{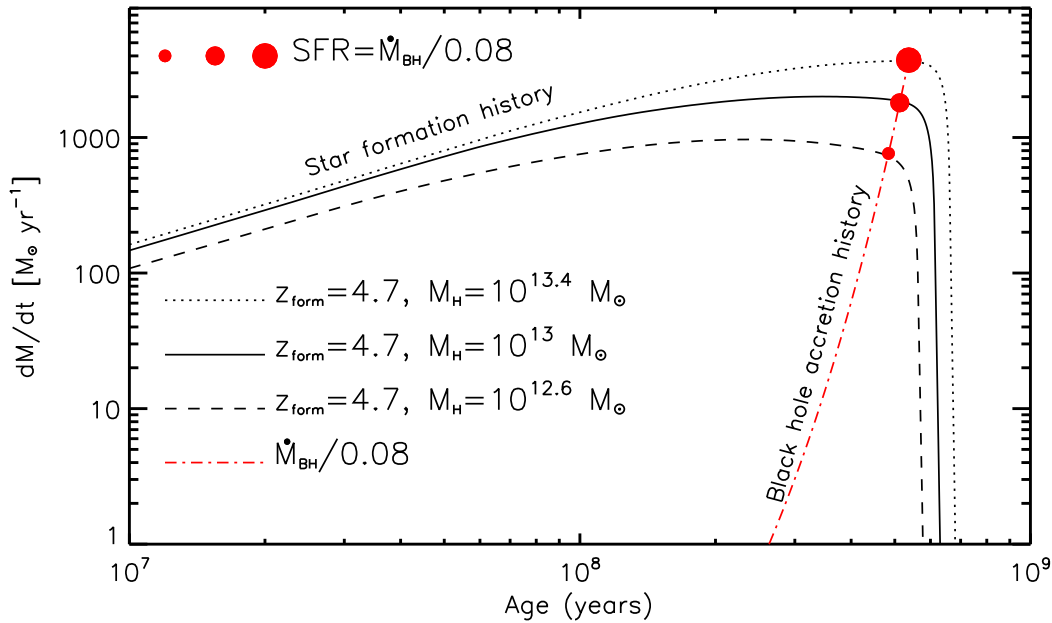}
\caption{Star formation histories and the black hole growth history predicted by the \citet{granato2004} model .
  SFHs are for dark matter halos virilized at formation redshift
  $z_{form}=4.7$ with masses $10^{12.60}$ (dashed line),$10^{13.00}$ (solid line) and $10^{13.40} M_\odot$ (dotted line), respectively.
  The dot-dashed line represents the scaled black hole growth history by multiplying the black hole growth rate, which is described
  in Equation \ref{equ:dmbh_dt}, with a factor of 1/0.08. The filled circles show the moment when SFR$=0.08\times\dot{M}_{BH}$ for a given SFH,
  as what we have observed in Hot DOGs. }\label{fig:sfh_model}
\end{figure*}

Here we attempt to examine if the observed \emph{extreme} properties of Hot DOGs can be predicted by
the model of galaxy formation and evolution. We employ a physical model for the coevolution of QSOs
and their hosts proposed by \citet{granato2004}, hereafter G04, to reproduce the observed properties of Hot DOGs.
In G04 model, star formation rate can be written as
\begin{equation}\label{equ:sfr_mcold}
{\rm SFR} = \int
\frac{\mathrm{d}M_{\mathrm{cold}}}{\max[t_{\mathrm{cool}},t_{\mathrm{dyn}}]}
\approx \frac {M_{\mathrm{cold}}}{t_{\star}}~,
\end{equation}
where $t_{\mathrm{cool}}$ and $t_{\mathrm{dyn}}$ are the cooling time and dynamical time, respectively.
$t_{\star}$ is the star formation timescale averaged over the mass distribution.
$M_{\mathrm{cold}}$ is the cold gas mass, which is dependent on the virilized dark matter halo
$M_H$ and the formation redshift $z_{form}$.
The black hole grows according to gas accretion at a given Eddington ratio $\lambda_{Edd}$:
\begin{equation}\label{equ:dmbh_dt}
\dot{M}_{BH}=\frac{\lambda_{\rm Edd}M_{BH}}{\tau_{Salp}}
\end{equation}
where $\tau_{Salp}$ is the Salpeter timescale \citep{salpeter1964}.
For the adopted value of $\eta$, where $\frac{\eta}{1-\eta}=0.1$, $\tau_{Salp}\sim 50$Myr.
We assume a seed black hole mass $M^{seed}_{BH}=10^3M_\odot$ and adopt $\lambda_{Edd}=1.5$.
In G04 model, star formation and black hole growth will be quenched by QSOs and SN{\ae} feedback
when star formation reaches its peak. More details on the model descriptions and analytical approximations
can be found in \citet{granato2004,lapi2006,lapi2014,mao2007,fan2008,fan2010,cai2013,cai2014}.

In Figure \ref{fig:sfh_model}, we plot the model predicted star formation histories (SFHs)
for dark matter halos virilized at formation redshift $z_{form}=4.7$ with halo masses $10^{12.60},10^{13.00}$ and $10^{13.40} M_\odot$, respectively.
We also plot the scaled black hole growth history by multiplying the black hole growth rate with a factor 1/0.08=12.5.
The filled circles represent the time when SFR$ =0.08\times\dot{M}_{BH}$ for a given SFH, as what we have observed
in Hot DOGs. The corresponding redshift at that time is about $z\sim3$, which is close to the median value of redshift distribution in Hot DOGs.
At the time marked by the filled circles, the model predicted SFRs are about $700,~2000$ and
$4500~M_\odot~ yr^{-1}$ respectively. And black hole masses vary from $\sim1.0\times10^9 M_\odot$ to $\sim1.0\times10^{10}M_\odot$.
As a comparison, SFRs in Hot DOGs span from 600 to $\sim6000~ M_\odot~ yr^{-1}$, with a
median value of $\sim2000~ M_\odot~ yr^{-1}$, and black hole masses in Hot DOGs span from
$\sim7.0\times10^8M_\odot$ to $\sim8.0\times10^{9}M_\odot$, assuming $\lambda_{Edd}=1.5$.
Both the predicted ranges of SFR and black hole mass are well consistent with the observations of Hot DOGs.
G04 model also predicts that around the peak of star formation history the intense star formation will be associated with
significant quantities of dust distributed in both AGN torus and hosts, which will bury the central accreting SMBH.
This is the probable case in Hot DOGs.

As seen in Figure \ref{fig:sfh_model}, the simple model can well reproduce the observed properties of Hot DOGs.
Several probable indications can be deduced from the comparison between the model and the observations:
(1) Hot DOGs may lie at or close to peaks of both star formation history and
black hole growth history. (2) Black hole grows exponentially while star formation has a relatively slow growth.
Black hole accretes most of its final mass during the last e-folding time.
As a consequence, there should be a dusty starburst dominated phase before the moment when
Hot DOGs have been observed. For instance, at $\sim 10^8$yr, SFR remains $\sim 1000 M_\odot yr^{-1}$, 
while $\dot{M}_{BH}$ would be smaller by $\sim3$ orders of magnitude (see Figure \ref{fig:sfh_model}).
These are exactly the observed properties of SMGs, which are known as dusty starbursts. 
Over peaks of both star formation and black hole accretion activities, QSOs feedback has been 
proposed to remove the remaining gas and dust, and then leave an optically bright QSO. 
The recent work by \citet{diaz2015} drawn the same conclusion, suggesting that one Hot DOG (W$2246-0256$) 
is near to bursting out the surrounding dust to become an optically bright QSO based on the study of 
spatially resolved ALMA [C\,{\sc ii}] observations. 
(3) Thus Hot DOGs may represent a transit phase during the evolution of massive galaxies, transforming from the dusty 
starburst dominated phase to the optically bright QSO phase. 
%At the peak of its mass growth, black hole looks to be outgrowing their host galaxies compared to the local spheroid mass-black hole mass relation.

\section{Summary}\label{sec:summary}

In this work, we select 22 submm-detected Hot DOGs with spectroscopic redshift. Their
observed IR SEDs have been constructed by combining \wise, \textit{Herschel} PACS and SPIRE, SCUBA-2 850$\mu$m data and
other available mm observations. We use a Bayesian SED analysis approach to decompose the observed
IR SEDs into two components: torus and cold dust. We use the CLUMPY model to describe torus emission and a
gray body to represent the cold dust emission related to star formation.
Our main results are summarized below.

\begin{enumerate}
  \item We compare the Bayesian evidences of Torus+GB with Torus models. We find that Torus+GB model has the higher
    Bayesian evidence for all Hot DOGs than Torus model. We also find that ${\rm ln} (\frac{{\rm ev_{TORUS+GB}}}{{\rm ev_{TORUS}}}$) $>10$
   (corresponding to odds of $> 20000:1$), which represents strong evidence in favor of Torus+GB model.
  \item Our submm-detected Hot DOGs are all hyperluminous IR galaxies (HyLIRGs, $L_{IR}\geq10^{13}L_\odot$) or extremely
  luminous IR galaxies (ELIRGs, $L_{IR}\geq10^{14}L_\odot$). Torus emission dominates the IR energy output. Cold dust emission is 
  averagely contributing no more than $\sim24\%$ of total IR luminosity, depending on the choice of torus models.
  \item We construct a median Hot DOG SED by taking the median value of 22 normalized rest-frame SEDs of Hot DOGs.
  The median SED is very steep at $1-5\mu$m and becomes rather flat at $\sim10-50~\mu$m,
  then sharply drops at $>100\mu$m. Hot DOGs have the highest luminosity ratio between mid-IR and submm at rest-frame
  compared to QSOs and starburst templates. The similarity between Hot DOGs and QSO SEDs at $\sim10-50~\mu$m
  suggests that the heating sources of Hot DOGs should be buried AGNs.
  \item Hot DOGs have high dust temperatures ($<T_{dust}> \sim72~$K) and high IR luminosities of cold dust $L^{cd}_{IR}$.
  Compared to high-$z$ SMGs and QSOs with similar $L^{cd}_{IR}$, Hot DOGs have the similar $T_{dust}-L_{IR}$ relation. We use the form
  $L_{IR}=4\pi R_e^2\sigma T^{4.35}$ to describe the expected $T_{dust}-L_{IR}$ relation of a gray body at $T_{dust}>35$K.
  We find that, at $L^{cd}_{IR}>10^{12}L_\odot$, the increase in IR luminosity is mostly due to the increase of dust temperature,
  rather than dust mass. Compared to SMGs at $z<4$, our Hot DOGs show higher dust temperature, but smaller $R_e$. Thus the increase
  in IR luminosities of our Hot DOGs relative to those of SMGs within similar redshift range should be dominated by the increase
  in dust temperature rather   than $R_e$. The increase of dust temperature could be due to the more intense radiation field caused
  by more intense starburst activity and/or buried AGN activity.
  \item The dust masses of Hot DOGs (Log M$_{dust}$ [M$_\odot$]) range from 7.5 to 8.6, with a median value of 7.9 which are about 0.4-0.5 dex
  lower than those of both SMGs and QSOs within similar redshift range. The lower dust masses in Hot DOGs is mainly due to the high dust
  temperature, as the dust mass estimation is strongly affected by $T_{dust}$ with  $M_{dust}\propto T_{dust}^{-2.3}$ at $T_{dust}\geq35$K.
  The lower dust masses in Hot DOGs will predict lower molecular gas masses. This is consistent with the non-detection
  of CO J$=4-3$ line by CARMA in \citet{wu2014}. We will use the deep CO line observations with ALMA 
  to examine this argument during ALMA Cycle 3 (PI: L. Fan).
  \item The dust covering factor of Hot DOGs spans from 0.4 to 1.0, which deviates from the trend that the dust covering factor decreases with
  increasing bolometric luminosity. Hot DOGs have heavily dust obscuration and high bolometric luminosity, which could
  have been missed in the previous UV/optical and X-ray AGN samples.
  \item We investigate the possible evolutionary path of Hot DOGs by employing a simple physical model. By comparing the model predictions and
  the observed properties, we suggest that Hot DOGs may lie at or close to both peaks of star formation 
  and black hole growth histories, and represent a transit phase during the evolution of massive galaxies, transforming from the dusty
  starburst dominated phase to the optically bright QSO phase.
\end{enumerate}

\begin{acknowledgements}
This work is supported by the National Natural Science Foundation of China (NSFC, Nos. 11203023, 11303084 and 11433005),
the Fundamental Research Funds for the Central Universities (WK3440000001). LF acknowledges the support by Qilu Young Researcher Project of Shandong
University. LF and KK acknowledge the Knut and Alice Wallenberg Foundation for support. YH 
thanks the support from the Western Light Youth Project. RN acknowledges support by
FONDECYT grant No. 3140436.

{\it Facilities:\/} \facility{\textit{WISE}}, \facility{\textit{Herschel} (PACS,SPIRE)}, \facility{\textit{JCMT} (SCUBA2)}.

\end{acknowledgements}

\appendix

\section{Bayesian approach for SED fitting}\label{bayesed}

BayeSED (\citeauthor{Han2012a} \citeyear{Han2012a}, \citeyear{HanY2014a})\footnote{https://bitbucket.org/hanyk/bayesed/} is designed to be a general purpose Bayesian SED fitting code, which means that it can be used to fit the multi-wavelength SEDs of galaxies with the combination of whatever SED models. Given any model SED library, which could be too large to be practically used, we first employ principal component analysis (PCA) to reduce the library dimensionality without sacrificing much accuracy.
Then, a supervised machine learning method, such as artificial neural network (ANN) algorithm, or K-Nearest Neighbors (KNN) searching, is employed to approximately generate the model SED at any position of the parameter space spanning by the model SED library.
So, by using these methods, the original SED model, which is given as a SED library, can be approximately and very efficiently evaluated at any position of its parameter space. Thanks to these efficient machine learning methods, we can break through the main bottleneck in Bayesian SED fitting \citep{Kauffmann2003a,SalimS2007a,daCunha2008a,Noll2009a,Walcher2011a}, which often require a very extensive sampling of a high-dimensional parameter space.

Similar to other Bayesian SED fitting codes \citep{asensio2009,Acquaviva2011a,Serra2011a,Johnson2013a}, we estimate the parameters of SED models by using the posterior probability distribution function (PDF) of parameters.
Instead of the more traditional Markov Chain Monte Carlo (MCMC) algorithm, we have employed the newly developed multimodal
nested sampling algorithm (MultiNest, Feroz et al. 2008,2009) to obtain the posterior PDF of parameters.
What makes MultiNest algorithm be different from MCMC algorithm is its ability to calculate the Bayesian evidence of a model and explore a more complicate parameter space with multiple posterior modes and pronounced (curving) degeneracies in moderately high dimensions.  This ability is crucial for a more reasonable analysis of very complicated multi-wavelength SEDs of galaxies.
When modeling the SEDs of galaxies \citep{Conroy2013e}, it is very common for us to have multiple physically reasonable choices.
So, it is very necessary to have a valid tool to discriminate between these possible choices.
The Bayesian evidence \citep{JeffreysH1998a,Jaynes2003a,Gregory2005a}, which quantitatively implements the principle of Occam's razor, can be employed as such a tool. According to the principle of Occam's razor, a model will not only be appreciated for a better explanation of observations but also be punished for more complexity. 

The Bayesian parameter estimation and model comparison with BayeSED have been demonstrated in \cite{Han2012a} for a sample of hyperluminous infrared galaxies by using the CLUMPY AGN torus model \citep{Nenkova2008a,Nenkova2008b} and the Starburst model of \cite{Siebenmorgen2007a}, and in \cite{HanY2014a} for a Ks-selected sample of galaxies in the COSMOS/UltraVISTA field by using stellar population synthesis models. In \cite{HanY2014a} we also presented an extensive test of the reliability of BayeSED code for SED fitting of galaxies. 
%Given these tests and thanks to those employed efficient machine learning methods and the parallelization of the code with Message Passing Interface, BayeSED code could be very suitable for IR SED analysis of Hot DOGs here.

\section{Model comparsion and model parameters}

The natural logarithm ${\rm ln(ev_{TORUS})}$, ${\rm ln(ev_{TORUS+GB}})$ of the Bayesian evidences for Torus and Torus+GB models and the natural logarithm of Bayes factor
${\rm ln} (\frac{{\rm ev_{TORUS+GB}}}{{\rm ev_{TORUS}}}$) have been presented in Table \ref{tbl-ev}.

In Table \ref{tbl-para} we also list the median values and 16\% and 84\% quartiles of seven free parameters ($Y,i,q,\sigma,N_0,\tau_V, T_{dust}$) and two derived quantities ($f_2,P_{type1}$) with the best-fitting TORUS+GB model.

\begin{table}
\centering
\caption{The Bayesian evidences of "TORUS" and "TORUS+GB" models\label{tbl-ev}}
%\begin{center}
\begin{tabular}{lccc}
\hline
\hline
Source & ln(ev$_{\rm TORUS}$) & ln(ev$_{\rm TORUS+GB}$) & ln ($\frac{{\rm ev_{TORUS+GB}}}{{\rm ev_{TORUS}}}$) \\
\hline

  W0126$-$0529  &  $-158.38\pm0.18$  &  $ -71.88\pm0.23$  &  $  86.50\pm0.41$  \\
  W0134$-$2922  &  $ -62.42\pm0.13$  &  $ -15.26\pm0.14$  &  $  47.16\pm0.28$  \\
  W0149+2350    &  $ -34.56\pm0.14$  &  $  -9.15\pm0.12$  &  $  25.41\pm0.26$  \\
  W0220+0137    &  $ -49.31\pm0.17$  &  $  -9.25\pm0.13$  &  $  40.06\pm0.30$  \\
  W0248+2705    &  $ -34.11\pm0.13$  &  $  -9.43\pm0.12$  &  $  24.68\pm0.24$  \\
  W0410$-$0913  &  $ -49.89\pm0.17$  &  $ -12.51\pm0.14$  &  $  37.37\pm0.31$  \\
  W0533$-$3401  &  $ -28.80\pm0.14$  &  $  -7.65\pm0.12$  &  $  21.15\pm0.26$  \\
  W0615$-$5716  &  $ -28.50\pm0.16$  &  $ -11.74\pm0.15$  &  $  16.76\pm0.31$  \\
  W0757+5113    &  $ -42.14\pm0.12$  &  $  -7.73\pm0.12$  &  $  34.41\pm0.24$  \\
  W0859+4823    &  $ -68.60\pm0.16$  &  $  -9.93\pm0.13$  &  $  58.68\pm0.29$  \\
  W1136+4236    &  $ -39.94\pm0.15$  &  $ -19.71\pm0.10$  &  $  20.23\pm0.25$  \\
  W1248$-$2154  &  $ -34.39\pm0.17$  &  $  -7.84\pm0.13$  &  $  26.55\pm0.29$  \\
  W1603+2745    &  $ -59.20\pm0.18$  &  $ -10.33\pm0.13$  &  $  48.87\pm0.30$  \\
  W1814+3412    &  $ -62.11\pm0.15$  &  $  -9.49\pm0.14$  &  $  52.62\pm0.29$  \\
  W1835+4355    &  $-191.90\pm0.19$  &  $ -13.28\pm0.15$  &  $ 178.62\pm0.34$  \\
  W2054+0207    &  $ -53.24\pm0.14$  &  $ -15.48\pm0.11$  &  $  37.77\pm0.25$  \\
  W2201+0226    &  $ -60.11\pm0.18$  &  $ -14.92\pm0.15$  &  $  45.19\pm0.33$  \\
  W2210$-$3507  &  $-112.54\pm0.16$  &  $ -18.67\pm0.14$  &  $  93.87\pm0.31$  \\
  W2216+0723    &  $ -25.41\pm0.14$  &  $  -7.15\pm0.11$  &  $  18.25\pm0.26$  \\
  W2238+2653    &  $ -90.39\pm0.16$  &  $ -11.86\pm0.14$  &  $  78.52\pm0.30$  \\
  W2246$-$0526  &  $ -32.61\pm0.14$  &  $ -22.29\pm0.16$  &  $  10.32\pm0.30$  \\
  W2305$-$0039  &  $-128.15\pm0.17$  &  $ -43.15\pm0.20$  &  $  85.01\pm0.36$  \\
\hline
\end{tabular}
\end{table}

\begin{table*}
\centering
\caption{The model parameters and derived quantities with the best-fitting TORUS+GB model\label{tbl-para}}
%\begin{center}
\begin{tabular}{lccccccccc}
\hline
\hline
Source & $N_0$ & $Y$ & $i$ & $q$ & $\sigma$ & $\tau_v$ & ${f_2}$ & ${P_{Type1}}$ & ${T_{dust}}$  \\
\hline
  W0126$-$0529  &   13.65$^{+  0.50}_{-  0.16}$  &    5.76$^{+  0.59}_{-  0.37}$  &   85.67$^{+  2.02}_{-  0.76}$  &    2.88$^{+  0.05}_{-  0.06}$  &   67.51$^{+  0.28}_{-  0.50}$  &   11.23$^{+  0.75}_{-  0.60}$  &    1.00$^{+  0.01}_{-  0.01}$  &    0.01$^{+  0.01}_{-  0.01}$  &   88.27$^{+  1.23}_{-  1.25}$  \\
  W0134$-$2922  &    5.16$^{+  1.09}_{-  0.87}$  &   48.51$^{+ 23.80}_{- 22.29}$  &   64.01$^{+ 13.10}_{- 20.53}$  &    2.86$^{+  0.07}_{-  0.07}$  &   52.80$^{+  8.16}_{-  9.16}$  &   17.34$^{+  5.29}_{-  3.77}$  &    0.89$^{+  0.05}_{-  0.10}$  &    0.03$^{+  0.14}_{-  0.02}$  &   74.94$^{+  5.67}_{-  5.36}$  \\
  W0149+2350  &    6.62$^{+  3.02}_{-  1.78}$  &   56.06$^{+ 20.51}_{- 22.60}$  &   48.02$^{+ 22.57}_{- 25.10}$  &    2.38$^{+  0.31}_{-  0.40}$  &   49.67$^{+  8.20}_{-  9.28}$  &   26.20$^{+ 10.89}_{-  7.65}$  &    0.91$^{+  0.05}_{-  0.11}$  &    0.05$^{+  0.29}_{-  0.04}$  &   75.44$^{+  6.39}_{-  6.17}$  \\
  W0220+0137  &    7.36$^{+  2.51}_{-  1.69}$  &   54.95$^{+ 20.46}_{- 19.64}$  &   59.50$^{+ 16.07}_{- 23.76}$  &    1.81$^{+  0.57}_{-  0.50}$  &   51.62$^{+  7.80}_{- 10.43}$  &   29.97$^{+ 11.17}_{-  8.51}$  &    0.94$^{+  0.04}_{-  0.11}$  &    0.02$^{+  0.10}_{-  0.01}$  &   80.14$^{+  6.28}_{-  6.45}$  \\
  W0248+2705  &    8.41$^{+  2.46}_{-  2.28}$  &   48.88$^{+ 25.31}_{- 26.83}$  &   42.23$^{+ 22.75}_{- 21.42}$  &    2.75$^{+  0.11}_{-  0.27}$  &   53.32$^{+  7.23}_{-  7.70}$  &   37.38$^{+ 22.25}_{- 13.04}$  &    0.93$^{+  0.05}_{-  0.10}$  &    0.03$^{+  0.36}_{-  0.03}$  &   66.74$^{+  5.31}_{-  5.27}$  \\
  W0410$-$0913  &    5.06$^{+  1.60}_{-  1.18}$  &   63.33$^{+ 16.46}_{- 19.40}$  &   54.89$^{+ 15.69}_{- 20.39}$  &    1.32$^{+  0.41}_{-  0.34}$  &   50.44$^{+  8.68}_{- 10.47}$  &   19.61$^{+  6.20}_{-  4.67}$  &    0.89$^{+  0.06}_{-  0.14}$  &    0.07$^{+  0.12}_{-  0.04}$  &   63.81$^{+  2.23}_{-  2.12}$  \\
  W0533$-$3401  &    6.07$^{+  3.14}_{-  1.97}$  &   47.20$^{+ 25.73}_{- 24.51}$  &   36.06$^{+ 25.13}_{- 19.03}$  &    2.45$^{+  0.29}_{-  0.35}$  &   41.27$^{+ 10.67}_{-  9.38}$  &   43.22$^{+ 16.94}_{- 12.49}$  &    0.83$^{+  0.07}_{-  0.17}$  &    0.39$^{+  0.60}_{-  0.33}$  &   76.81$^{+  3.68}_{-  3.32}$  \\
  W0615$-$5716  &    5.22$^{+  0.87}_{-  0.66}$  &   72.78$^{+ 11.92}_{- 14.59}$  &   74.52$^{+  7.20}_{-  8.10}$  &    1.18$^{+  0.29}_{-  0.18}$  &   54.80$^{+  7.33}_{-  8.83}$  &   17.18$^{+  3.99}_{-  3.59}$  &    0.91$^{+  0.04}_{-  0.11}$  &    0.01$^{+  0.01}_{-  0.01}$  &   67.19$^{+ 13.62}_{- 13.01}$  \\
  W0757+5113  &    8.92$^{+  2.66}_{-  2.17}$  &   44.57$^{+ 26.46}_{- 25.02}$  &   53.49$^{+ 17.43}_{- 24.07}$  &    2.68$^{+  0.16}_{-  0.23}$  &   54.12$^{+  6.67}_{-  8.21}$  &   35.19$^{+ 14.24}_{- 10.07}$  &    0.96$^{+  0.03}_{-  0.06}$  &    0.01$^{+  0.16}_{-  0.01}$  &   45.43$^{+  3.80}_{-  3.15}$  \\
  W0859+4823  &    7.18$^{+  2.42}_{-  2.02}$  &   40.70$^{+ 27.54}_{- 19.22}$  &   46.42$^{+ 22.35}_{- 22.76}$  &    2.53$^{+  0.24}_{-  0.27}$  &   51.05$^{+  7.56}_{-  8.26}$  &   25.38$^{+  9.17}_{-  6.80}$  &    0.93$^{+  0.04}_{-  0.09}$  &    0.05$^{+  0.29}_{-  0.04}$  &   59.11$^{+  1.87}_{-  1.98}$  \\
  W1136+4236  &    8.68$^{+  2.88}_{-  2.85}$  &   44.54$^{+ 27.21}_{- 19.17}$  &   27.46$^{+ 28.70}_{- 15.22}$  &    2.27$^{+  0.35}_{-  0.42}$  &   41.54$^{+  8.67}_{-  8.22}$  &  139.25$^{+ 63.22}_{- 60.30}$  &    0.88$^{+  0.06}_{-  0.12}$  &    0.27$^{+  0.67}_{-  0.25}$  &   59.84$^{+  2.65}_{-  3.00}$  \\
  W1248$-$2154  &    6.90$^{+  3.42}_{-  2.04}$  &   41.06$^{+ 25.37}_{- 19.61}$  &   33.40$^{+ 25.49}_{- 18.73}$  &    2.48$^{+  0.31}_{-  0.40}$  &   46.53$^{+  8.75}_{-  6.75}$  &   40.77$^{+ 12.56}_{- 11.03}$  &    0.90$^{+  0.03}_{-  0.09}$  &    0.18$^{+  0.51}_{-  0.15}$  &   81.60$^{+  6.24}_{-  5.48}$  \\
  W1603+2745  &    4.55$^{+  2.45}_{-  1.59}$  &   50.87$^{+ 22.36}_{- 23.17}$  &   49.41$^{+ 19.23}_{- 24.03}$  &    2.78$^{+  0.11}_{-  0.25}$  &   43.09$^{+ 12.13}_{- 12.08}$  &   26.13$^{+ 17.67}_{-  8.09}$  &    0.75$^{+  0.13}_{-  0.18}$  &    0.29$^{+  0.63}_{-  0.25}$  &   66.39$^{+  2.90}_{-  2.89}$  \\
  W1814+3412  &   10.07$^{+  2.34}_{-  2.41}$  &   40.11$^{+ 27.10}_{- 19.79}$  &   49.91$^{+ 17.99}_{- 22.25}$  &    2.72$^{+  0.15}_{-  0.25}$  &   57.30$^{+  5.30}_{-  6.11}$  &   29.62$^{+ 11.45}_{-  8.25}$  &    0.97$^{+  0.02}_{-  0.04}$  &    0.00$^{+  0.05}_{-  0.00}$  &   69.97$^{+  4.52}_{-  4.22}$  \\
  W1835+4355  &    8.49$^{+  2.19}_{-  2.06}$  &   40.86$^{+ 24.20}_{- 19.21}$  &   53.99$^{+ 17.96}_{- 21.84}$  &    2.83$^{+  0.08}_{-  0.10}$  &   53.13$^{+  6.72}_{-  7.78}$  &   22.03$^{+  7.35}_{-  5.61}$  &    0.95$^{+  0.04}_{-  0.10}$  &    0.01$^{+  0.10}_{-  0.01}$  &   66.98$^{+  2.12}_{-  2.67}$  \\
  W2054+0207  &    3.84$^{+  3.05}_{-  1.42}$  &   47.14$^{+ 23.68}_{- 22.77}$  &   43.21$^{+ 21.35}_{- 19.09}$  &    2.73$^{+  0.14}_{-  0.27}$  &   35.37$^{+ 14.10}_{- 10.06}$  &   38.14$^{+ 28.83}_{- 15.48}$  &    0.65$^{+  0.15}_{-  0.15}$  &    0.56$^{+  0.43}_{-  0.43}$  &   77.37$^{+  5.84}_{-  6.03}$  \\
  W2201+0226  &    5.32$^{+  1.35}_{-  1.01}$  &   42.00$^{+ 23.14}_{- 19.88}$  &   61.16$^{+ 14.74}_{- 20.73}$  &    2.85$^{+  0.07}_{-  0.07}$  &   51.19$^{+  8.85}_{- 10.43}$  &   18.74$^{+  5.71}_{-  4.70}$  &    0.87$^{+  0.06}_{-  0.14}$  &    0.04$^{+  0.24}_{-  0.03}$  &   73.92$^{+  1.65}_{-  1.64}$  \\
  W2210$-$3507  &    9.25$^{+  2.31}_{-  1.84}$  &   57.34$^{+ 20.08}_{- 24.04}$  &   58.46$^{+ 16.24}_{- 23.46}$  &    2.55$^{+  0.22}_{-  0.26}$  &   55.12$^{+  6.65}_{-  7.52}$  &   25.45$^{+  9.26}_{-  6.30}$  &    0.96$^{+  0.02}_{-  0.06}$  &    0.00$^{+  0.04}_{-  0.00}$  &   52.60$^{+  1.51}_{-  1.34}$  \\
  W2216+0723  &    8.98$^{+  2.87}_{-  2.74}$  &   49.49$^{+ 23.23}_{- 23.74}$  &   20.09$^{+ 20.56}_{- 10.88}$  &    2.49$^{+  0.23}_{-  0.22}$  &   41.21$^{+  8.49}_{-  8.55}$  &  169.48$^{+ 51.39}_{- 53.83}$  &    0.88$^{+  0.05}_{-  0.13}$  &    0.39$^{+  0.58}_{-  0.36}$  &   73.22$^{+  3.62}_{-  3.38}$  \\
  W2238+2653  &    8.25$^{+  2.43}_{-  1.57}$  &   45.46$^{+ 22.67}_{- 22.87}$  &   43.49$^{+ 21.06}_{- 20.41}$  &    2.66$^{+  0.18}_{-  0.26}$  &   56.55$^{+  5.63}_{-  6.41}$  &   49.45$^{+ 16.13}_{- 11.50}$  &    0.96$^{+  0.02}_{-  0.05}$  &    0.01$^{+  0.13}_{-  0.01}$  &   70.06$^{+  2.27}_{-  2.59}$  \\
  W2246$-$0526  &    6.86$^{+  1.45}_{-  1.12}$  &   75.70$^{+ 12.99}_{- 13.10}$  &   16.97$^{+ 11.05}_{-  7.65}$  &    0.95$^{+  0.13}_{-  0.16}$  &   49.89$^{+  7.93}_{-  8.17}$  &   14.06$^{+  3.18}_{-  1.98}$  &    0.92$^{+  0.04}_{-  0.11}$  &    0.40$^{+  0.29}_{-  0.16}$  &   94.62$^{+  2.79}_{-  3.96}$  \\
  W2305$-$0039  &    6.68$^{+  0.63}_{-  0.47}$  &    6.48$^{+  1.54}_{-  0.77}$  &   79.01$^{+  5.89}_{-  7.51}$  &    2.89$^{+  0.05}_{-  0.06}$  &   60.65$^{+  4.06}_{-  2.83}$  &   12.44$^{+  1.75}_{-  1.28}$  &    0.96$^{+  0.01}_{-  0.01}$  &    0.00$^{+  0.00}_{-  0.00}$  &   81.23$^{+  2.49}_{-  2.52}$  \\

\hline
\end{tabular}
\end{table*}

\section{SED fitting}

\begin{figure*}
\plotone{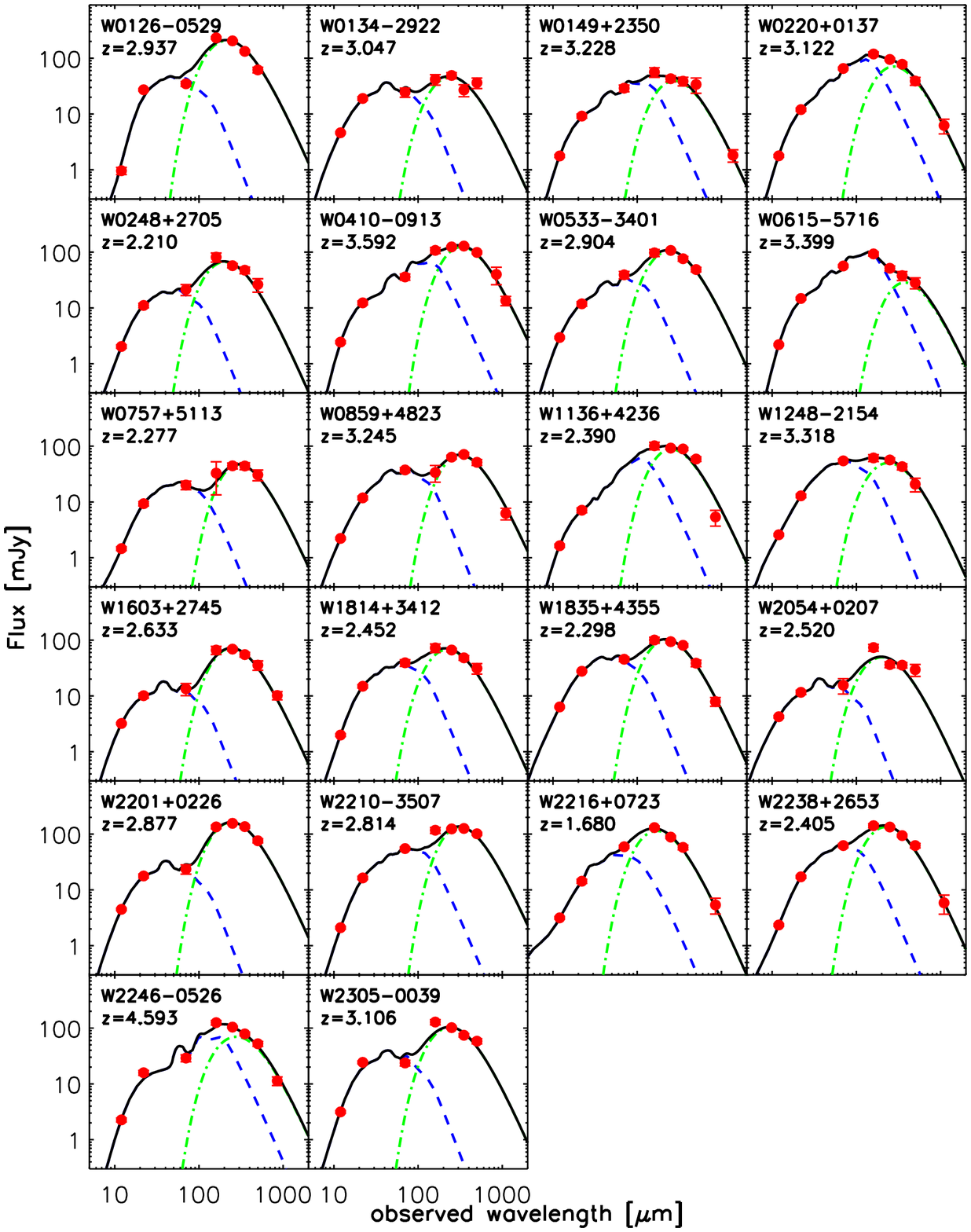}
\caption{Best-fit (or MAP) model SEDs with Torus+GB model for 22 Hot DOGs in our sample. The filled circles represent the observed data points.
The dashed, dot-dashed and solid lines are the same as in Figure \ref{fig:mod_comp}.}\label{fig:sed}
\end{figure*}

In Figure \ref{fig:sed}, we plotted the best-fit (or the maximum a posteriori, i.e., MAP) model SEDs adopting the Torus+GB model
for 22 Hot DOGs in our sample. In all cases, the torus component has a dominant contribution to the SED at observed wavelength
shorter than 100$\mu$m, which corresponds to $<25\mu$m at rest frame roughly for the Hot DOGs at $z\sim3$, while
the gray body component has a significant contribution at $>$100$\mu$m.

\end{CJK*}

\end{document}